\documentclass[twocolumn, trackchanges]{aastex631}
\usepackage[normalem]{ulem}
\newcommand{\Msun}{\mbox{$M_{\odot}$}}

\newcommand{\Mi}{\mbox{$M_{\rm i}$}}

\usepackage{booktabs}
\usepackage{diagbox}
\usepackage{tabularx}
\usepackage{pifont}
\newcommand{\xmark}{\ding{55}}%
\usepackage{multirow}

\begin{document}

%\title{Understanding the Initial-Final Mass Relation: Influence of Convective Overshooting and Mass Loss}
\title{The Impact of the Third Dredge-up and Mass Loss in Shaping the Initial-Final Mass Relation of White Dwarfs}

\author[0000-0002-3867-9966]{Francesco Addari}
\affiliation{Scuola Internazionale Superiore di Studi Avanzati, Via Bonomea, 265, I-34136, Trieste}
\affiliation{Osservatorio Astronomico di Padova-INAF, Vicolo dell'Osservatorio 5, I-35122
Padova, Italy}
\author[0000-0002-9137-0773]{Paola Marigo}
\affiliation{Dipartimento di Fisica e Astronomia Galileo Galilei, Università degli studi di Padova,\\ Vicolo dell’Osservatorio 3, I-35122 Padova, Italy}
\author[0000-0002-7922-8440]{Alessandro Bressan}
\affiliation{Scuola Internazionale Superiore di Studi Avanzati, Via Bonomea, 265, I-34136, Trieste}
\author[0000-0002-6213-6988]{Guglielmo Costa}
\affiliation{Univ Lyon, Univ Lyon1, Ens de Lyon, CNRS, Centre de Recherche Astrophysique de Lyon UMR5574, \\F-69230 Saint-Genis-Laval, France}
\author[0000-0001-5231-0631]{Kendall Shepherd}
\affiliation{Scuola Internazionale Superiore di Studi Avanzati, Via Bonomea, 265, I-34136, Trieste}
\affiliation{Osservatorio Astronomico di Padova-INAF, Vicolo dell'Osservatorio 5, I-35122
Padova, Italy}
\author[0000-0002-8691-4940]{Guglielmo Volpato}
\affiliation{Dipartimento di Fisica e Astronomia Galileo Galilei, Università degli studi di Padova,\\ Vicolo dell’Osservatorio 3, I-35122 Padova, Italy}
\affiliation{Osservatorio Astronomico di Padova-INAF, Vicolo dell'Osservatorio 5, I-35122
Padova, Italy}



\begin{abstract}
The initial-final mass relation (IFMR) plays a crucial role in understanding stellar structure and evolution by linking a star's initial mass to the mass of the resulting white dwarf.
This study explores the IFMR in the initial mass range $0.8 \leq M_\mathrm{ini} / M_\odot \leq 4$ using full \texttt{PARSEC} evolutionary calculations supplemented with \texttt{COLIBRI} computations to complete the ejection of the envelope and obtain the final core mass.
Recent works have shown that the supposed monotonicity of the IFMR is interrupted by a kink in the initial mass range $M_\mathrm{ini} \approx 1.65-2.10 M_\odot$, due to the interaction between recurrent dredge-up episodes and stellar winds in carbon stars evolving on the thermally-pulsing asymptotic giant branch phase.
To reproduce the IFMR non-monotonic behavior we investigate the role of convective overshooting efficiency applied to the base of the convective envelope ($f_\mathrm{env}$) and to the borders of the pulse-driven convective zone ($f_\mathrm{pdcz}$), as well as its interplay with mass loss.
We compare our models to observational data and find that $f_\mathrm{env}$ must vary with initial mass in order to accurately reproduce the IFMR's observed kink and slopes.
We find some degeneracy between the overshooting parameters when only the IFMR information is used.
Nonetheless, this analysis provides valuable insights into the internal mixing processes during the TP-AGB phase.

\end{abstract}

\keywords{asymptotic giant branch stars, carbon stars, stellar wind, white dwarfs, convective overshooting}


\section{Introduction} \label{sec:intro}
The Thermally-Pulsing Asymptotic Giant Branch (TP-AGB) phase is the final evolutionary stage of low- and intermediate-mass stars ($0.8 \lesssim \Mi/\Msun \lesssim 7-8$), which ends when the envelope is ejected by stellar winds in the interstellar medium and the bare central core cools as a carbon-oxygen (C-O) white dwarf \citep{karakas_dawes_2014, herwig_evolution_2005}.

The evolution of the TP-AGB phase is heavily influenced by processes that are difficult to model from first principles, such as turbulent convection, stellar winds, long-period variability, and various dredge-up episodes (collectively designated as third dredge-up, or TDU).
Furthermore, these processes interact with one another and do not occur smoothly during the TP-AGB evolution, but they may vary greatly in characteristics and efficiency over a thermal pulse cycle (TPC).

We recall that the TDU is responsible for the formation of carbon stars, characterized by a photospheric C/O ratio larger than one. In fact, at the stage of the maximum luminosity produced by the thermal instabilities of the He-burning shell, the bottom of the convective envelope may stretch inside the region involved in the thermal pulse nucleosynthesis, bringing newly synthesized carbon produced by the triple-alpha reaction to the surface \citep{herwig_evolution_2005}. 
 Unfortunately, the TDU is highly dependent on the physics prescriptions and numerical treatments \citep{frost_numerical_1996, mowlavi_third_1999} of the stellar evolution code, resulting in a wide heterogeneity of TP-AGB models and results \citep{straniero_evolution_1997, herwig_evolution_2000, stancliffe_third_2005, weiss_new_2009, cristallo_evolution_2011, marigo_evolution_2013, karakas_helium_2014, ventura_gas_2018}. 
 
One method for constraining the efficiency of the TDU is to reproduce the carbon star luminosity functions in various galaxies with different age-metallicity relations and known star formation histories
\citep{groenewegen1993, marigo_third_1999, marigo_girardi_07, Pastorelli_etal_19, Pastorelli_etal_20}.

The semi-empirical initial final mass relation (IFMR) of C-O white dwarfs provides another approach to calibrating the TDU in Milky Way carbon stars, as demonstrated in a few studies \citep{Kalirai_etal_14, marigo_carbon_2020, Marigo_universe_22}.

Following the exploration in these works, this research focuses on the semi-empirical IFMR, with the goals of reproducing semi-empirical behavior using recent data and retrieving information on the TDU using full self-consistent TP-AGB stellar models.
To recover the semi-empirical IFMR, we will specifically investigate the role of mass loss and convective overshooting applied to the base of the convective envelope and the borders of the pulse-driven convective zone (PDCZ).

The IFMR is an important tool for understanding stellar evolution because it provides insight into the processes that occur during the star's lifetime, in particular setting constraints on the amount of mass lost by stellar winds \citep{Marigo_iau_13, Kalirai_etal_14, Marigo_universe_22, Marigo_iau_22}.

In general, the IFMR predicts that more massive stars will produce more massive remnants. Over the years improvements on the semi-empirical IFMR have been achieved thanks to new observations and refined treatments in stellar evolution codes. \citep[and references therein]{weidemann_revision_2000,Williams_etal_04,kalirai_initial-final_2008, kalirai_masses_2009, salaris_semi-empirical_2009, Williams_etal_09, Kalirai_etal_14,cummings_white_2018, cummings_novel_2019}.

Recently, with the addition of new WDs data belonging to open clusters with ages of 1.5-2.5 Gyr, \citet{marigo_carbon_2020} found a kink in the IFMR at about $\Mi\ \simeq 1.65-2.10 M_\odot$, that suddenly interrupts the commonly assumed monotonic behavior. 
Surprisingly, the white dwarfs at the peak, which are all members of the open cluster NGC 7789, reach masses of $\simeq 0.70-0.74 \Msun$, which have previously been associated with stars with $\Mi \simeq 3 \Msun$.

The IFMR kink is interpreted as the signature of the lowest-mass stars in the Milky Way that evolved into carbon stars during the TP-AGB phase. 
According to \cite{marigo_carbon_2020}, these carbon stars are expected to have undergone shallow third dredge-up events, resulting in low photospheric C/O and modest carbon excess with respect to oxygen, C-O. Under these conditions, carbonaceous dust grains cannot condense in sufficient quantities to cause a strong wind, so the TP-AGB lifetime is prolonged and the core mass can grow more than usually predicted. Theoretically, the key to this behavior is the use of mass-loss prescriptions for carbon stars that are dependent on carbon excess, which are based on state-of-the-art dynamical models \citep{mattsson_dust_2010, eriksson_synthetic_2014, bladh_carbon_2019}.

 An independent study of AGB stars in Galactic open clusters using Gaia EDR3 recently confirmed these findings \citep{marigo_fresh_2021}. In the initial mass range of the kink, they found carbon stars with dust-free spectra and irregular small-amplitude pulsations (implying very low mass loss as estimated from spectral energy distribution fitting) with current core masses of $\simeq 0.67-0.7\, M_{\odot}$, consistent with WD masses (see Figure 8 of that paper).

 The paper is structured as follows. Section \ref{sec:inputphys} presents and discusses the input physics and technical details of our \texttt{PARSEC} models. In Section \ref{sec:evol}, we recall the evolutionary properties of low- and intermediate-mass stars with references to our tracks. Section \ref{sec:final_core} describes how to estimate the final mass of the white dwarf and the shape of the IFMR. Our concluding remarks end the paper in Section \ref{sec:conclusion}. 

\section{Input Physics} \label{sec:inputphys}
The code and main input physics used in \texttt{PARSEC V2.0} is described thoroughly in \cite{bressan_parsec_2012}, \cite{tang_new_2014}, \cite{chen_improving_2014}, \cite{fu_new_2018}, \cite{costa_multiple_2019, costa_mixing_2019} and \cite{nguyen_parsec_2022}. Here we summarise the relevant points for the low- and intermediate-mass stars treated in this work.

Our sets of TP-AGB tracks have solar-like metallicity $Z = 0.014$, with a solar-scaled chemical composition from \cite{caffau_solar_2011}. The helium content is given by $Y = Z \cdot \Delta Y / \Delta Z + Y_\mathrm{p} = 0.273$ with $Y_\mathrm{p} = 0.2485$ \citep{komatsu_cosmic_2011} and $\Delta Y / \Delta Z = 1.78$ \citep{bressan_parsec_2012}. Equation of state tables are calculated using the \texttt{FREEEOS} code developed by A. W. Irwin\footnote{http://freeeos.sourceforge.net/}. Nuclear reaction rates includes a total of 72 reactions and tracks 32 isotopes: $^{1}$H, D, $^{3}$He, $^{4}$He, $^{7}$Li, $^{7}$Be, $^{12}$C, $^{13}$C, $^{14}$N, $^{15}$N, $^{16}$O, $^{17}$O, $^{18}$O, $^{19}$F, $^{20}$Ne, $^{21}$Ne, $^{22}$Ne, $^{23}$Na, $^{24}$Mg, $^{25}$Mg, $^{26}$Mg, $^{27}$Al, $^{28}$Al, $^{29}$Si, $^{30}$S, $^{31}$Ar, $^{40}$Ca, $^{44}$Ti, $^{48}$Cr, $^{52}$Fe, $^{56}$Ni, $^{60}$Zn.
The primary energy-generating nuclear reactions are all included; a list with references can be found in Table 1 of \cite{fu_new_2018} and \cite{costa_formation_2021}. We now detail the physics inputs and go over all of the updates for dealing with the TP-AGB phase.

Opacity tables in the low-temperature regime ($\log(T/{\rm K}) < 4.2$) are calculated with \texttt{{\AE}SOPUS} \citep{marigo_low-temperature_2009,marigo_low-temperature_2022} and have been updated to track the variation of carbon, nitrogen and oxygen abundances due to consecutive dredge-up events and hot-bottom burning. In the high-temperature regime, $4.2 \le \log(T/{\rm K}) \le 8.7$, we use
the opacity tables provided by the Opacity Project At Livermore \citep[\texttt{OPAL};][]{OPAL_96}.
Conductive opacities are incorporated following \citet{Itoh_etal_08}.

Nuclear reaction network and mixing are solved together with diffusive equations \citep{costa_mixing_2019}.
Regions unstable to convection are defined by the Schwarzchild criterion. 
The diffusion coefficient within these zones is derived from the mixing-length theory framework \citep{MLT_58}
and we adopt a mixing length parameter $\alpha=1.74$ according to the calibration of the standard solar model \citep{bressan_parsec_2012}.
$\alpha$ is fixed across the whole evolution.
At each border of all unstable zones we include convective overshooting.
These regions are considered radiative.
Moreover, core overshooting is treated as a ballistic process \citep{bressan_mass_1981}, while the diffusion coefficient for the convective envelope and pulse-driven convective zone overshooting is calculated with the scheme proposed by \citet{herwig_evolution_2000}:
\begin{equation} \label{eq:exponential_overshooting}
    D(r) = D_0 \exp \left( -2\frac{ \left| r-r_0 \right| }{f_\mathrm{ov} H_\mathrm{p}} \right) \quad \quad r_0 = r_\mathrm{cnv} \pm f_\mathrm{0,ov} H_\mathrm{p}
\end{equation}
where $r_\mathrm{cnv}$ is the radial coordinate of the convective border according to the Schwarzchild criterion and $f_\mathrm{0,ov} H_\mathrm{p}$ is the distance inside the convective region at which overshooting begins to be applied and where the diffusion coefficient is equal to $D_0$. The minimum value of the diffusion coefficient is set to $D_\mathrm{min} = 10^3$ 
$\mathrm{cm}^2$ $\mathrm{s}^{-1}$, below which no mixing is allowed. For simplicity we assume $f_\mathrm{ov} = 2 f_\mathrm{0,ov}$ \citep{choi_mesa_2016}. We define $f_\mathrm{env}$ and $f_\mathrm{pdcz}$ respectively as the overshooting parameters at the bottom of the convective envelope and in the PDCZ.
In the latter, convective overshooting is applied on both borders of the instability region.
Previous studies have shown indications of overshooting in the PDCZ \citep{herwig_evolution_2000, wagstaff_impact_2020} but there is no general agreement yet.
A first principle analysis indicates that the inertia of the convective eddies in the PDCZ is not enough to make them overcome the Schwarzchild border \citep{lattanzio_overshoot_2017}. This may imply that there must be another mechanism able to mix over the canonical border of the PDCZ. Exponential overshooting is a simple and effective prescription that can overcome the canonical convective border while lacking a more physically sound scheme.

As described by \cite{bressan_parsec_2012} and \cite{nguyen_parsec_2022}, we apply core and envelope overshooting depending on the initial mass of the track.
We define $M_\mathrm{O1}$ as the minimum initial mass that has a convective core during the main sequence.
For $M_\mathrm{ini} < M_\mathrm{O1}$ we do not apply core overshooting, and $f_\mathrm{env} = f_\mathrm{env,min} = 0.033$; for $M_\mathrm{ini} > M_\mathrm{O2} = M_\mathrm{O1} + 0.3 M_\odot$ we apply maximum core and envelope overshooting, as $\lambda_\mathrm{ov} = \lambda_\mathrm{ov,max} = 0.5$ and $f_\mathrm{env} = f_\mathrm{env,max} = 0.047$; for $M_\mathrm{O1} \le M_i \le M_\mathrm{O2}$ both parameters ($\lambda_\mathrm{ov}$ and $f_\mathrm{env}$) scale linearly with the initial mass.
We will refer to this set of models as $f_\mathrm{env}=0.047^\star$ or "fiducial" value.
$f_\mathrm{env,min}$ and $f_\mathrm{env,max}$ are the corresponding values of $\Lambda_\mathrm{ov,min} = 0.5$ and $\Lambda_\mathrm{ov,max} = 0.7$ calibrated on the RGB bump and blue loop width for the previous envelope overshooting prescription \citep{alongi_effects_1991, christensen-dalsgaard_more_2011, bressan_uncertainties_2015, fu_new_2018}.
The conversion from the previous to the exponential overshooting prescriptions has been performed by matching the position and width of the RGB bump in the HR diagram for low-mass and intermediate-mass.
During central He-burning phase the efficiency of core overshooting is set to $\lambda_\mathrm{ov,max}$ for every initial mass, which gives consistent horizontal branch and AGB lifetimes with R$_2$ ratio observed in globular clusters \citep{bressan_evolution_1986, constantino_treatment_2016}. We do not suppress the occurrence of breathing pulses \citep{sweigart_semiconvection_1973, castellani_helium-burning_1985}, but their efficiency decreases with the inclusion of non-local core overshooting \citep{bressan_evolution_1986}.
We do not include the effects of rotation or magnetic fields, and the only extra-mixing comes from overshooting.

Mass loss is treated as follows: we assume no stellar wind up to the end of the main sequence. Once the star leaves the main sequence (MS) and proceeds through the red giant branch we use the Reimers' law with $\eta_\mathrm{R} = 0.2$ \citep{reimers_circumstellar_1975}. 
After the core He-burning is complete and early AGB begins, the stellar wind is described as a two-stage process. The entire formulation of mass loss on the AGB is taken from \citet{marigo_carbon_2020}.
For luminosities below the tip of the red-giant branch, we assume the Alfven waves-driven wind by \cite{cranmer_testing_2011}. Above this threshold, while the photospheric C/O remains below unity,  the mass loss is caused by pulsations and radiation pressure on silicate dust grains \citep{bloecker_stellar_1995} with $\eta_\mathrm{B} = 0.01$. As the star surface becomes carbon-rich, with a low carbon excess with respect to oxygen, it enters a phase with little or no amounts of dust, and the stellar wind is described by pulsating models of almost dust-free atmospheres \citep{winters_systematic_2000}. Finally, when the carbon excess is large enough ( $8.2 \lesssim \mathrm{(C-O)}_\mathrm{min} \lesssim 9.2$ depending on current mass, luminosity and effective temperature) to form significant amounts of carbonaceous dust we use the state-of-the-art dynamical atmosphere models by \cite{mattsson_dust_2010}, \cite{ eriksson_synthetic_2014} and \cite{ bladh_carbon_2019}. 

Lastly, we developed, tested, and used a shell-shifting-like treatment for accelerating the calculation during quiescent interpulses (details in Appendix \ref{sec:shellshift}).

We calculated a total of 439 tracks distributed on 24 sets. The pre-TP-AGB evolution of each track in every set has been calculated with fiducial envelope overshooting $f_\mathrm{env} = 0.047^\star$, as explained above. Details are discussed in the following section. Then, each set is identified by a couple of values ($f_\mathrm{env}$, $f_\mathrm{pdcz}$), which refer only to the TP-AGB phase. Our sets span the values $0.047^\star \leq f_\mathrm{env} \leq 0.160$ and $0 \leq f_\mathrm{pdcz} \leq 0.064$, as summarized in Table \ref{tab:sets}.
\begin{table}[htbp]
\centering
\small
\begin{tabularx}{\columnwidth}{@{\extracolsep{\fill}}cccccccc}
\toprule \toprule
\multirow{2}{*}{$f_\mathrm{pdcz}$} & \multicolumn{7}{c}{$f_\mathrm{env}$} \\ \cmidrule(lr){2-8}
& 0.047* & 0.056 & 0.064 & 0.096 & 0.128 & 0.144 & 0.160 \\ \cmidrule(r){1-1} 
0.000 & \textcolor[HTML]{32CB00}{\checkmark} & \textcolor[HTML]{CB0000}{\text{$\times$}} & \textcolor[HTML]{32CB00}{\checkmark} & \textcolor[HTML]{CB0000}{\text{$\times$}} & \textcolor[HTML]{32CB00}{\checkmark} & \textcolor[HTML]{CB0000}{\text{$\times$}} & \textcolor[HTML]{CB0000}{\text{$\times$}}\\
0.001 &	\textcolor[HTML]{32CB00}{\checkmark} & \textcolor[HTML]{32CB00}{\checkmark} & \textcolor[HTML]{32CB00}{\checkmark} & \textcolor[HTML]{32CB00}{\checkmark} & \textcolor[HTML]{32CB00}{\checkmark} & \textcolor[HTML]{32CB00}{\checkmark} & \textcolor[HTML]{32CB00}{\checkmark} \\
0.002 & \textcolor[HTML]{32CB00}{\checkmark} & \textcolor[HTML]{CB0000}{\text{$\times$}} & \textcolor[HTML]{32CB00}{\checkmark} & \textcolor[HTML]{CB0000}{\text{$\times$}} & \textcolor[HTML]{32CB00}{\checkmark} & \textcolor[HTML]{CB0000}{\text{$\times$}} & \textcolor[HTML]{CB0000}{\text{$\times$}} \\
0.004 & \textcolor[HTML]{32CB00}{\checkmark} & \textcolor[HTML]{CB0000}{\text{$\times$}} & \textcolor[HTML]{32CB00}{\checkmark} & \textcolor[HTML]{CB0000}{\text{$\times$}} & \textcolor[HTML]{32CB00}{\checkmark} & \textcolor[HTML]{CB0000}{\text{$\times$}} & \textcolor[HTML]{CB0000}{\text{$\times$}}\\
0.008 & \textcolor[HTML]{32CB00}{\checkmark} & \textcolor[HTML]{CB0000}{\text{$\times$}} & \textcolor[HTML]{32CB00}{\checkmark} & \textcolor[HTML]{CB0000}{\text{$\times$}} & \textcolor[HTML]{32CB00}{\checkmark} & \textcolor[HTML]{CB0000}{\text{$\times$}} & \textcolor[HTML]{CB0000}{\text{$\times$}}\\
0.016 & \textcolor[HTML]{32CB00}{\checkmark} & \textcolor[HTML]{CB0000}{\text{$\times$}} & \textcolor[HTML]{32CB00}{\checkmark} & \textcolor[HTML]{CB0000}{\text{$\times$}} & \textcolor[HTML]{32CB00}{\checkmark} & \textcolor[HTML]{CB0000}{\text{$\times$}} & \textcolor[HTML]{CB0000}{\text{$\times$}}\\
0.032 & \textcolor[HTML]{32CB00}{\checkmark} & \textcolor[HTML]{CB0000}{\text{$\times$}} & \textcolor[HTML]{CB0000}{\text{$\times$}} & \textcolor[HTML]{CB0000}{\text{$\times$}} & \textcolor[HTML]{CB0000}{\text{$\times$}} & \textcolor[HTML]{CB0000}{\text{$\times$}} & \textcolor[HTML]{CB0000}{\text{$\times$}}\\
0.064 & \textcolor[HTML]{32CB00}{\checkmark} & \textcolor[HTML]{CB0000}{\text{$\times$}} & \textcolor[HTML]{CB0000}{\text{$\times$}} & \textcolor[HTML]{CB0000}{\text{$\times$}} & \textcolor[HTML]{CB0000}{\text{$\times$}} & \textcolor[HTML]{CB0000}{\text{$\times$}} & \textcolor[HTML]{CB0000}{\text{$\times$}}\\ \toprule \bottomrule
\end{tabularx}
\caption{Sampled values of ($f_\mathrm{env}$, $f_\mathrm{pdcz}$). Green checkmarks correspond to the calculated sets of tracks, and red crosses to combinations of overshooting parameters that have not been explored.} \label{tab:sets}
\end{table}
The tracks span the range of $0.8 \leq \Mi/\Msun\leq 4$ (only a few sets extend to $\Mi\sim 5 M_\odot$) and the mass step is $0.05 \leq \Delta M_\mathrm{ini} / M_\odot \leq 0.3$ depending on the initial mass. Tracks close to and within the IFMR kink have a finer grid than stars outside.

The evolution is calculated from the Pre-Main Sequence (PMS) to the furthest point in the TP-AGB (see Sect. \ref{sec:final_core}). As in \citet{bressan_parsec_2012} the code interrupts when a star approaches the He-flash and then it is restarted from a proper Zero Age He-burning model (ZAHB), with mass corresponding to the total mass left at the RGB tip. The transition is so rapid that we can assume the star does not lose mass. We calculated proper ZAHB models for our fiducial set $f_\mathrm{env} = 0.047$*. Overshooting in the PDCZ clearly does not matter at this stage. We determined that stars do experience He-flash for $M_\mathrm{ini} \leq 1.85 M_\odot$ with an accuracy of $0.05 M_\odot$. We also estimated our $M_\mathrm{O1} = 1.22 M_\odot$, which corresponds to the H-burning core being convective and from which we start to apply core overshooting and the minimal value for $f_\mathrm{env}$ in our fiducial set.

\section{Evolutionary properties} \label{sec:evol}
In this section, we highlight the evolutionary features of our models, with a specific focus on the core mass evolution. The definition of the core mass varies depending on the evolutionary stage. The evolution before the TP-AGB phase of \texttt{PARSEC} tracks have been thoroughly described recently by \citet{nguyen_parsec_2022}. Prior to the TP-AGB phase, $M_\mathrm{core}$ is equal to the mass of the hydrogen-exhausted core. In order to include the possibility that white dwarfs may retain a thin hydrogen atmosphere on top of the He-intershell and CO core \citep{saumon_current_2022}, we take
\begin{equation}
    M_\mathrm{core} = m( X = 0.5 X_\mathrm{surf} )
\end{equation}
as the definition of the core mass in TP-AGB. $m$ is a generic mass coordinate, $X$ and $X_\mathrm{surf}$ are the hydrogen abundance in any mesh point and in the surface respectively. We also mark the beginning of TP-AGB where the thickness of the He-intershell falls below 0.1 $M_\odot$ \citep{dotter_mesa_2016}. 

\subsection{Pre-TP-AGB Evolution}
\begin{figure}[!t]
    \centering
    \includegraphics[width=0.45\textwidth]{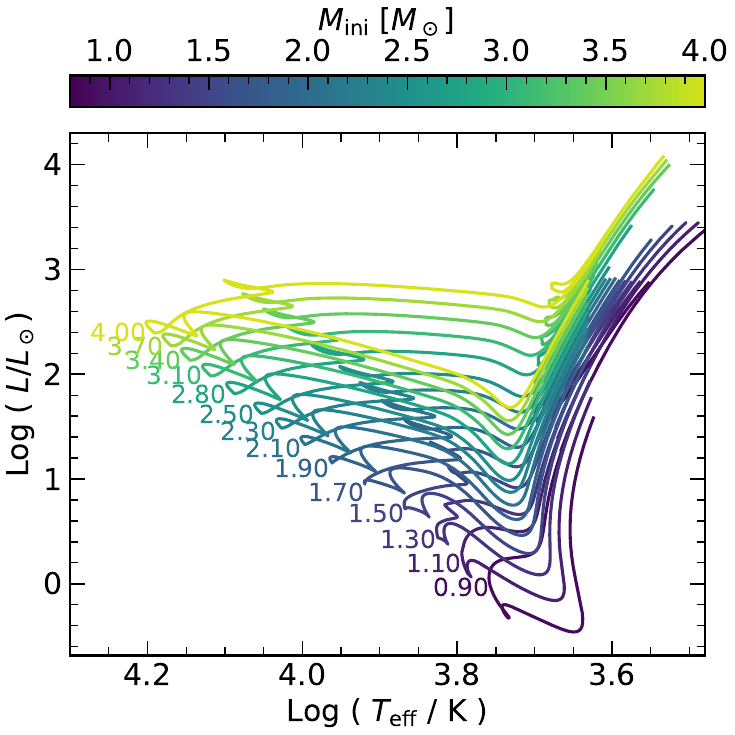}
    \caption{HR diagram for the fiducial value of envelope overshooting and no PDCZ overshooting, from the PMS up to the onset of the TP-AGB phase. A subset is shown for clarity.}
    \label{fig:colored_hr}
\end{figure} 
Fig.~\ref{fig:colored_hr} shows the evolution in the HR diagram of a sub-sample of the spanned mass range until the beginning of the TP-AGB phase. With our assumptions of $f_\mathrm{env}$, this part is common for all sets of tracks.

Before TP-AGB, the core is built up by the ashes of (core and shell) hydrogen and helium burning. The growth can be limited by the first and second dredge-up events (respectively FDU and SDU). These are directly related to the efficiency of envelope overshooting. 

\begin{figure}
    \centering
    \includegraphics[height=0.35\textheight]{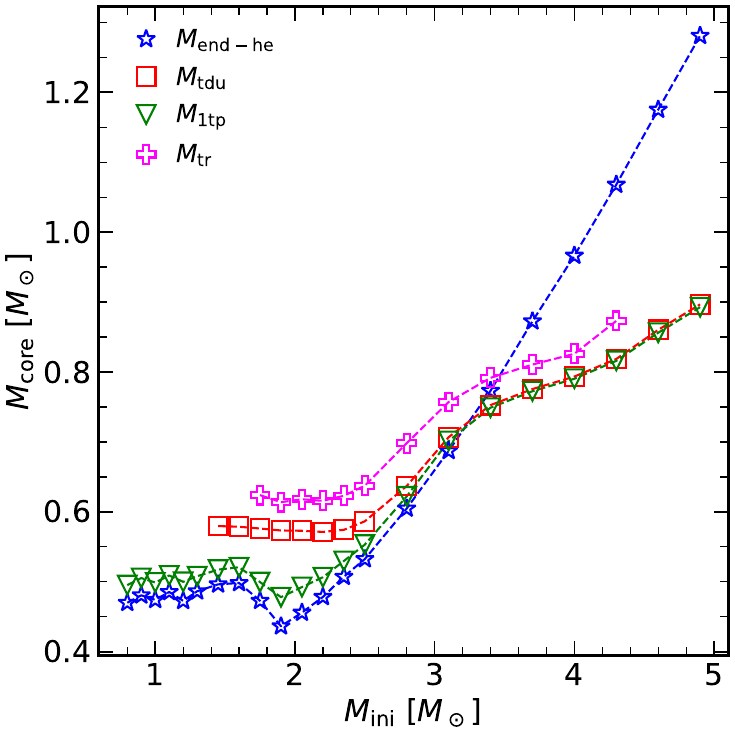} % Replace with your image file name
    \caption{Core mass at different stages of the evolution for fiducial envelope overshooting and $f_\mathrm{pdcz}=0.001$. $M_\mathrm{end-he}$ (blue stars) refers to the end of central He-burning, $M_\mathrm{1tp}$ (green triangles) to the first thermal pulse, $M_\mathrm{tdu}$ (red squares) the first occurrence of TDU and finally $M_\mathrm{tr}$ (pink plus) refers to the core mass at which the star moves from M- to C-type.}
    \label{fig:massthresholds}
\end{figure}

Our study focuses on the core mass evolution and the final mass of the white dwarf left at the end. Our choice for setting the same $f_\mathrm{env}$ for all initial masses before TP-AGB fixes the penetration of the envelope during the first and second dredge-up. Observations of the RGB bump constrain $f_\mathrm{env}$ during the FDU \citep{alongi_effects_1991, fu_new_2018}; later during core He burning, $f_\mathrm{env}$ affects the extensions and position of blue loops \citep{tang_new_2014}. However, there are no direct constraints for the overshooting efficiency during SDU. Therefore we want to make sure the mass of the core and surface C/O ratio are minimally impacted by the SDU experienced by tracks with different values of $f_\mathrm{env}$.
The SDU eventually affects stars with $M_\mathrm{ini} \gtrsim 3.5 M_\odot$ \citep{karakas_dawes_2014}, thus we calculated few intermediate-mass stars tracks setting the $f_\mathrm{env}$ value from the PMS, different from the fiducial $f_\mathrm{env} = 0.047^\star$..
In Figure~\ref{fig:massthresholds} various mass thresholds are shown with definitions in the figure caption. Where $M_\mathrm{end-he} > M_\mathrm{1tp}$ reveals which tracks experience the SDU, in agreement with \citet{karakas_dawes_2014}.
Table \ref{tab:sdu} shows the main properties at the on-set of the TP-AGB phase of two models with $M_\mathrm{ini} = 3.7 M_\odot$ and $ M_\mathrm{ini} = 4 M_\odot$, computed with different $f_\mathrm{env}$ from the PMS.
In particular, the core mass after the second dredge-up and the carbon-to-oxygen ratio are the two main properties that will shape the following evolution of the track.
These quantities differ by less than 2\% in models with different $f_\mathrm{env}$ (from the PMS), and there is no clear trend with increasing overshooting penetration.
Thus, we conclude that these differences may be led by purely numerical features, which are intrinsic in the mixing treatment.

Our test confirmed that the SDU is not appreciably affected by the choice of envelope overshooting, at least up to $M_\mathrm{ini} \simeq 4 M_\odot$. Then, we modify the value of $f_\mathrm{env}$ only during the TP-AGB, ensuring consistency with previous calibrations. PDCZ is only present during a TP, thus the previous evolution is insensitive to $f_\mathrm{pdcz}$.

\begin{table}
%\scriptsize
\centering
\begin{tabularx}{0.85\columnwidth}{@{\extracolsep{\fill}}cccc}
  \toprule \toprule
        \multicolumn{4}{c}{$M_\mathrm{ini} = 3.70 M_\odot$} \\
        \midrule
    $f_\mathrm{env}$ &    0.047 &    0.096 &    0.144 \\ \midrule
   $M_\mathrm{core}$ &    0.791 &    0.778 &    0.782 \\ 
             $X_\mathrm{surf}$ & 6.538E-01 & 6.522E-01 & 6.530E-01 \\ 
             $Y_\mathrm{surf}$ & 3.321E-01 & 3.337E-01 & 3.330E-01 \\ 
         $X_\mathrm{C12,surf}$ & 1.456E-03 & 1.447E-03 & 1.445E-03 \\ 
         $X_\mathrm{O16,surf}$ & 5.137E-03 & 5.106E-03 & 5.098E-03 \\ 
       C/O$_\mathrm{surf}$ & 3.964E-01 & 3.965E-01 & 3.964E-01 \\
       \midrule
        \multicolumn{4}{c}{$M_\mathrm{ini} = 4.00 M_\odot$} \\
        \midrule
    $f_\mathrm{env}$ &    0.047 &    0.096 &    0.144 \\ \midrule
   $M_\mathrm{core}$ &    0.810 &    0.802 &    0.810 \\ 
             $X_\mathrm{surf}$ & 6.398E-01 & 6.389E-01 & 6.388E-01 \\ 
             $Y_\mathrm{surf}$ & 3.462E-01 & 3.471E-01 & 3.471E-01 \\ 
         $X_\mathrm{C12,surf}$ & 1.423E-03 & 1.417E-03 & 1.413E-03 \\ 
         $X_\mathrm{O16,surf}$ & 5.011E-03 & 4.985E-03 & 4.974E-03 \\ 
       C/O$_\mathrm{surf}$ & 3.975E-01 & 3.979E-01 & 3.977E-01 \\ 
  \bottomrule \bottomrule
    \end{tabularx}
    \caption{Track properties at the onset of TP-AGB phase, after SDU. $M_\mathrm{core}$ is in solar units. }
    \label{tab:sdu}
\end{table}

\subsection{TP-AGB evolution}

During the TP-AGB phase, the star undergoes recurring thermal pulses driven by the thermal instability of the geometrically thin helium shell. Following each pulse, the star may experience the TDU \citep{herwig_evolution_2005, karakas_dawes_2014}. This event reduces the core mass and transports helium-burning ashes to the surface. The efficiency of the third dredge-up is quantified by the parameter $\lambda$, defined as:

\begin{equation}
\lambda = \frac{\Delta M_\mathrm{dup}}{\Delta M_\mathrm{core}}
\end{equation}

where $\Delta M_\mathrm{dup}$ represents the decrease of the core mass caused by the penetration of the envelope after the $i$-th thermal pulse. $\Delta M_\mathrm{core}$ is equal to the growth of the core mass between the ($i-1$)-th and $i$-th pulses. It's important to acknowledge that $\lambda$ is highly sensitive to both the numerical details and the physical inputs of the model \citep{frost_numerical_1996, mowlavi_third_1999}. To ensure accurate predictions, maintaining consistent numerical prescriptions is crucial when calibrating the physical parameters to prevent potential systematic errors. We also recall that $\lambda$ is mainly dependent on the core mass and the envelope mass at fixed metallicity \citep{straniero_low-mass_2003}. These pieces of information serve as our reference point for the subsequent analysis of our TP-AGB tracks.
\begin{figure*}[bhtp]
    \centering
    \includegraphics[height=0.35\textheight]{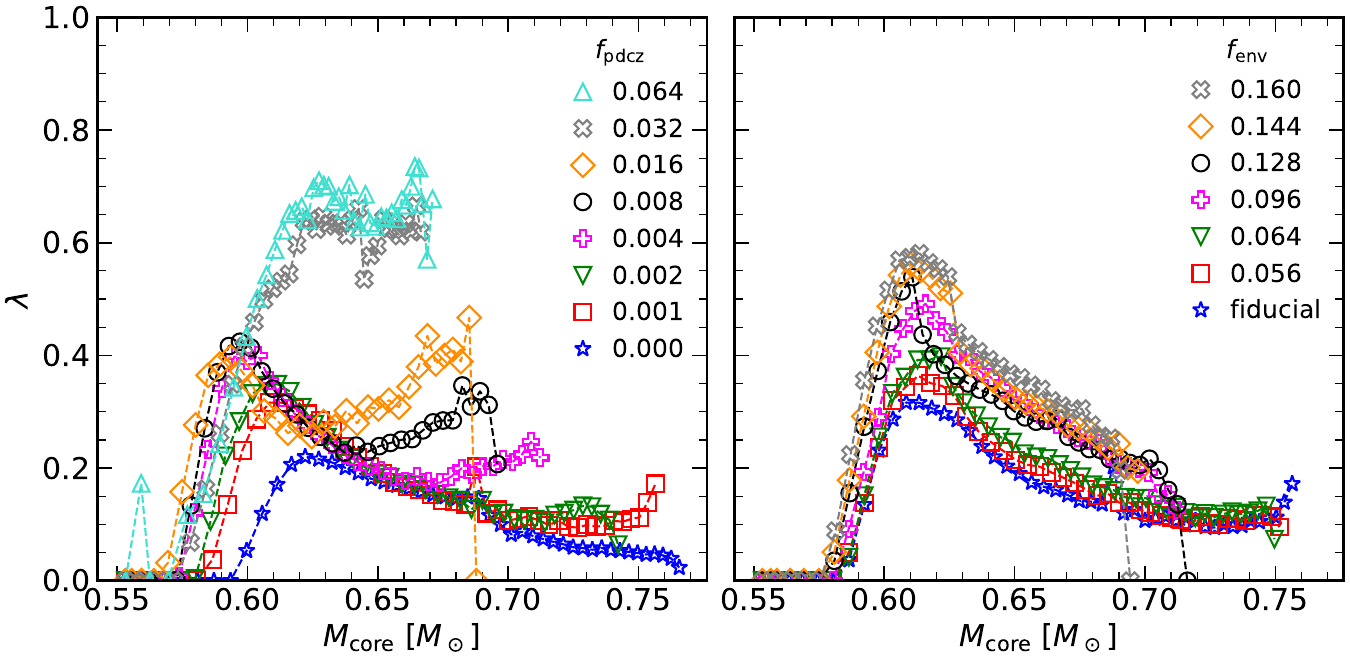}
    \caption{Evolution of $\lambda$ as a function of the core mass. Both panels show a $M_\mathrm{ini} = 2.50 M_\odot$ model. Left panel shows all $(0.047^\star, f_\mathrm{pdcz})$ models. Right panels shows  $(f_\mathrm{env}, 0.001)$ models. Each symbol corresponds to a thermal pulse.} \label{fig:tdu_lambda}
\end{figure*}
Figure~\ref{fig:tdu_lambda} illustrates the evolution of $\lambda$ relative to $M_\mathrm{core}$ for a $M_\mathrm{ini} = 2.5 M_\odot$ star for every set. 
Generally, increasing envelope overshooting leads to a more efficient dredge-up. That is a standard behavior and what is usually found in literature, as increased overshooting depth destabilizes deeper mass shells.
Instead $\lambda$-curves show a unique pattern at varying $f_\mathrm{pdcz}$. Again the TDU efficiency increases with the overshooting parameter (in this case PDCZ), in agreement with \citet{herwig_evolution_2000}. Nevertheless, the shape of the $\lambda$-curves at varying $f_\mathrm{pdcz}$ displays a prominent double maximum, instead of the usual bell-like profile \citep{straniero_evolution_1997, straniero_low-mass_2003, cristallo_evolution_2011, Marigo_universe_22}. We realized that the second rise of $\lambda$ with $M_\mathrm{core}$ begins roughly at $M_\mathrm{tr}$, at which the mass-loss rate changes from a Bl\"ocker wind to the dust-free pulsation-only driven regime. Thus, we argue that the track is moved back to a region of the ($M_\mathrm{core}$, $M_\mathrm{env}$) plane where TDU is favored \citep{straniero_low-mass_2003}.

\begin{figure}
    \centering
    \includegraphics[height=0.35\textheight]{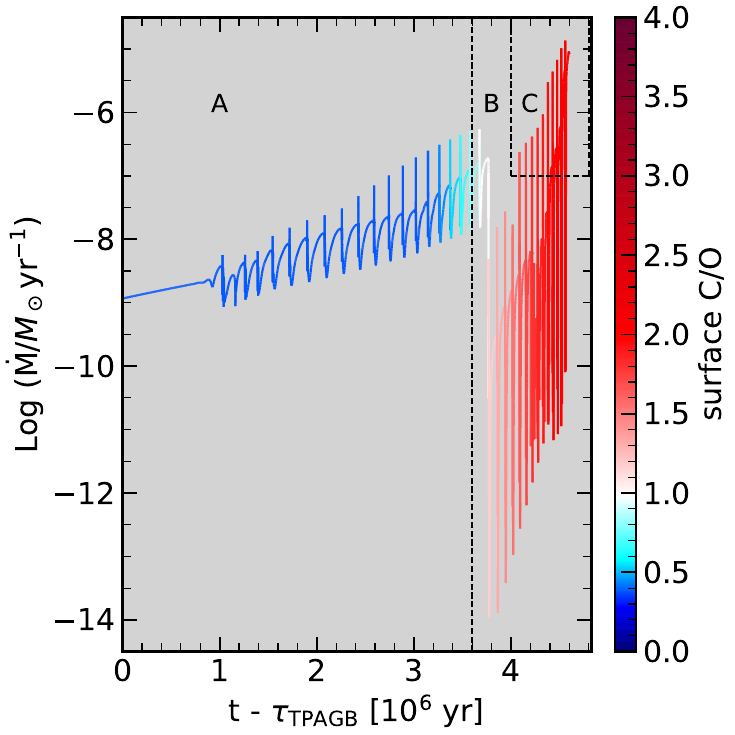} % Replace with your image file name
    \caption{Mass-loss evolution of $M_\mathrm{ini}=2.5 M_\odot$ with $f_\mathrm{env}=0.128$ and $f_\mathrm{pdcz}=0.001$. $\tau_\mathrm{TPAGB} = 1.350$ Gyr is the time spent before the beginning of TP-AGB. A, B and C mark the main three regimes of the mass-loss. A: wind for O-rich stars \citep{bloecker_stellar_1995}; B: dust-free pulsation-driven wind \citep{marigo_carbon_2020}; C: carbon dust-driven wind \citep{mattsson_dust_2010, bladh_carbon_2019}.}
    \label{fig:masslossevol}
\end{figure}
\begin{figure*}[thbp]
    \centering
    \includegraphics[width=0.95\textwidth]{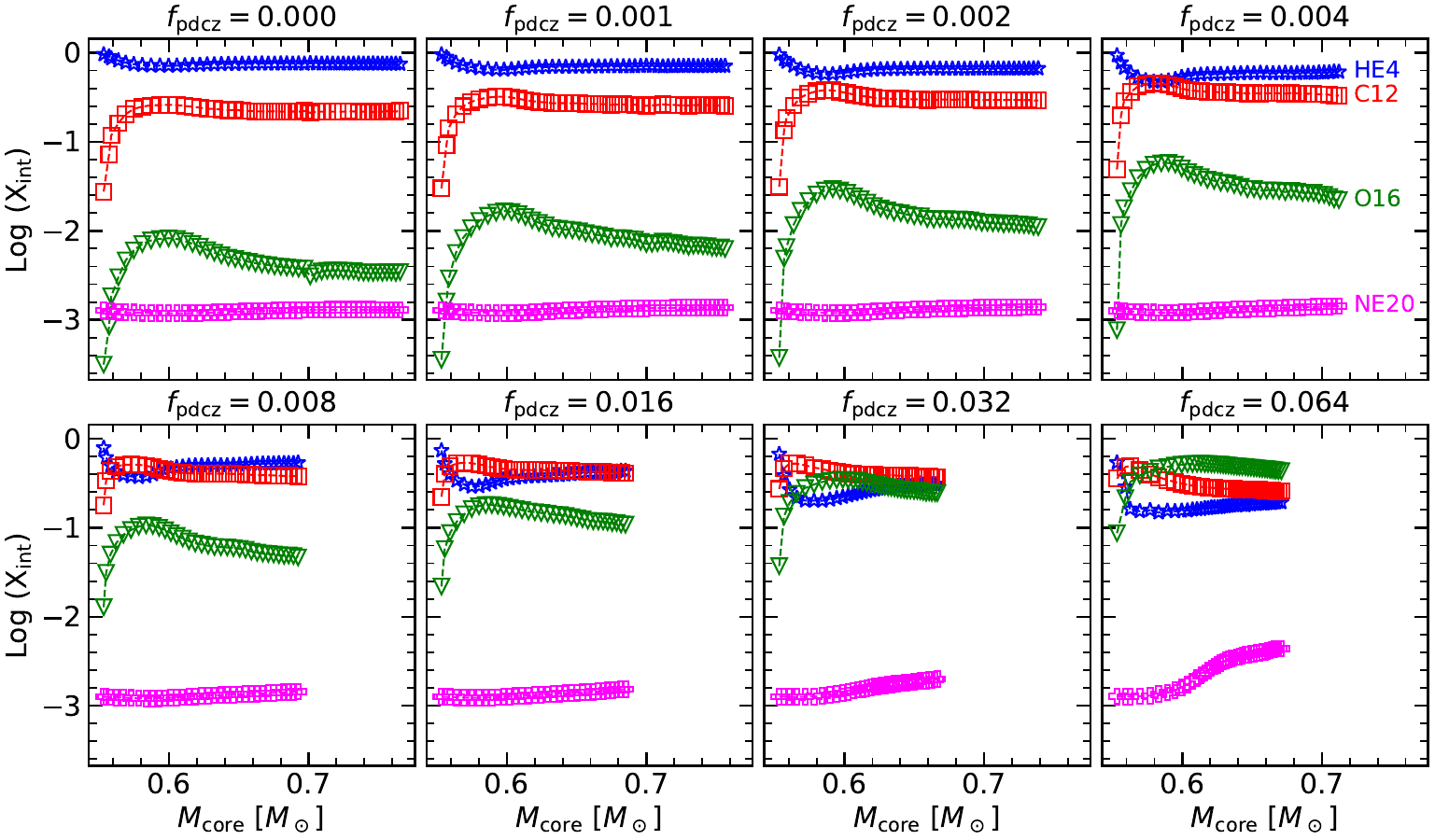}
    \caption{Intershell abundances (in log scale) of $^{4}$He, $^{12}$C,  $^{16}$O, and $^{20}$Ne after every calculated pulse for every ($0.047^\star$,$f_\mathrm{pdcz}$) couple and $M_\mathrm{ini} = 2.50 M_\odot$.}
    \label{fig:intershell}
\end{figure*}

It is important to highlight that $\lambda$ is not sufficient to determine completely the star's evolution. Instead, the interplay of the growth of the core mass and the composition of the dredge-up material is pivotal. Fig.~\ref{fig:masslossevol} shows an example of mass-loss rate evolution with time of a $M=2.50M_\odot$ star from the set (0.128,0.001). Changing $f_\mathrm{env}$ or $f_\mathrm{pdcz}$ does affect the evolution, as it changes the transition to carbon star in the first place, but it also modifies the duration of the dust-free phase with small mass-loss rates. It is not easy to predict how much the total TP-AGB lifetime changes and thus the effect on the core mass, but it is clear that the two overshooting parameters are mildly degenerate. $f_\mathrm{env}$ only impacts the TDU efficiency (as shown in Figure~\ref{fig:tdu_lambda}), while $f_\mathrm{pdcz}$ changes the intershell composition too, as shown in Figure~\ref{fig:intershell}. In general, deeper PDCZ overshooting will increase the carbon abundance in the intershell but it increases oxygen abundance too, decreasing the overall C/O$_\mathrm{intershell}$ ratio of the material that is being brought to the surface.

\begin{figure*}[thbp]
    \centering
    \includegraphics[width=0.96\textwidth]{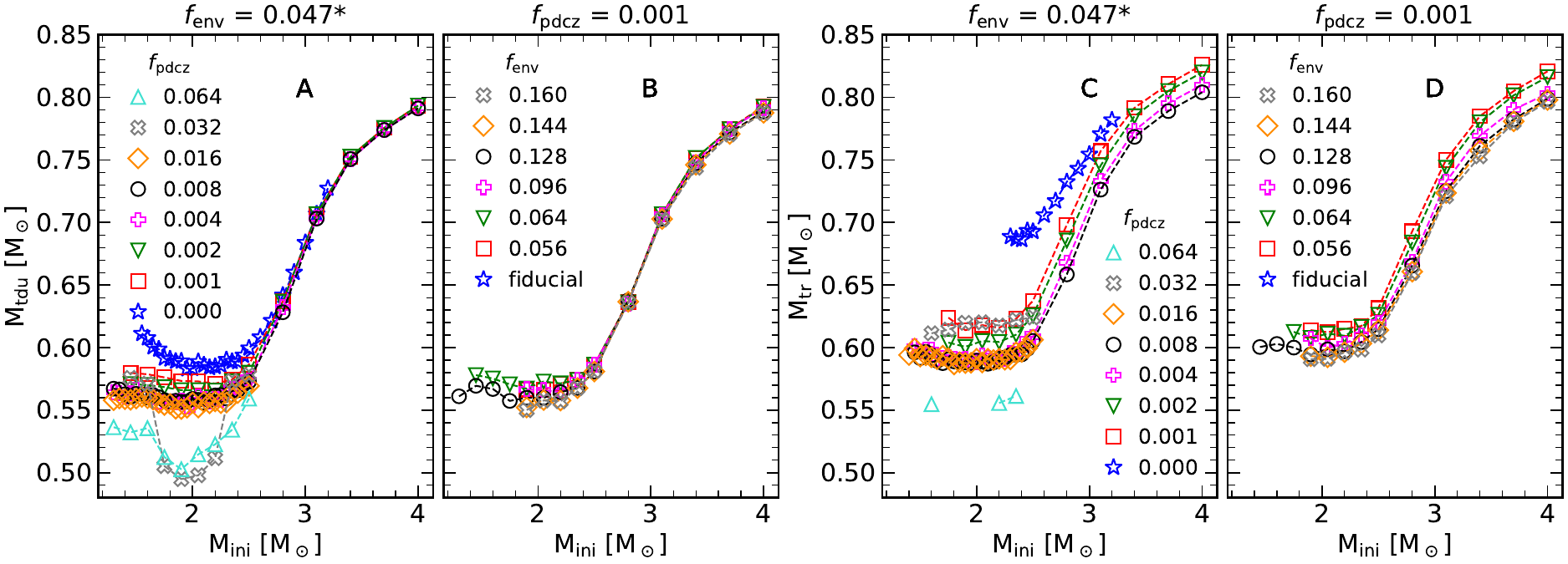}
    \caption{Core mass at the first occurrence of TDU (A and B panels) and core mass when $\mathrm{C}/\mathrm{O}_\mathrm{surf} > 1$ (C and D).  $f_\mathrm{pdcz} \geq 0.016$ models are limited to $M_\mathrm{ini} \leq 2.5 M_\odot$. $f_\mathrm{pdcz} = 0.000$ set is limited to $M_\mathrm{ini} \leq 3.2 M_\odot$. A and C panels: ($0.047^\star$,$f_\mathrm{pdcz}$) models. B and D panels: ($f_\mathrm{env}$,$0.001$) models. $f_\mathrm{env}=0.056$, $0.096$, $0.144$, $0.160$ are limited to $M_\mathrm{ini} \geq 1.9 M_\odot$.} \label{fig:first_tdu}
\end{figure*}

After analyzing the effect of $f_\mathrm{env}$ and $f_\mathrm{pdcz}$ on $\lambda$, we are interested in the impact on the mass threshold previously presented in Figure~\ref{fig:massthresholds}. Clearly, $M_\mathrm{end-he}$ and $M_\mathrm{1tp}$ are not affected due to our assumptions. Figure~\ref{fig:first_tdu} shows how the extension of the overshooting region (in the envelope and PDCZ) affects $M_\mathrm{tdu}$ and $M_\mathrm{tr}$. The effect of $f_\mathrm{env}$ is limited, as both $M_\mathrm{tdu}$ and $M_\mathrm{tr}$ change by less than 0.05$M_\odot$ in the low mass range. At $M_\mathrm{ini} \gtrsim 2.8 M_\odot$ the $M_\mathrm{tdu}$ does not change (in both cases) because it takes place at the first thermal pulse, which is independent of the overshooting parameters. On the other hand $f_\mathrm{pdcz}$ has a greater impact on $M_\mathrm{tdu}$ and $M_\mathrm{tr}$.
The first occurrence of TDU is significantly lowered in core mass at increasing $f_\mathrm{pdcz}$, especially for oxygen-rich intershells. Furthermore, TDU occurs for even lower initial masses, from $M_\mathrm{ini} \simeq 1.6 M_\odot$ at $f_\mathrm{pdcz} = 0.000$ to $M_\mathrm{ini} \simeq 1.3 M_\odot$ for $f_\mathrm{pdcz} \geq 0.004$. That is even more evident in looking at the transition core mass $M_\mathrm{tr}$ from M-type to C-type. A small value for $f_\mathrm{pdcz}$ is enough to make $M_\mathrm{tr}$ drop from $\sim 0.7 M_\odot$ to $\sim 0.63 M_\odot$. Again, it significantly lowers the threshold for the first (initial) mass becoming C-type from $M_\mathrm{ini} \simeq 2.3 M_\odot$ for no PDCZ-overshooting to $M_\mathrm{ini} \simeq 1.4$ $M_\odot$ at $f_\mathrm{pdcz} \simeq 0.008-0.016$. However, for $f_\mathrm{pdcz} > 0.016$, the trend reverses due to a combination of higher TDU efficiency and a C-deficient intershell, leading to delayed or non-existent transitions. For the extreme $f_\mathrm{pdcz} = 0.064$, only three tracks become C-type, but they remain in this phase for a brief period before reverting to M-type due to dredge-ups from an oxygen-rich intershell.

We have investigated the main effects of the two overshooting parameters and their interplay with mass loss on the quantities that shape the IFMR. These considerations are fundamental in shaping the final form of the Initial-Final Mass Relation (IFMR), which is discussed in the following section.

\section{Estimate of the final core mass} \label{sec:final_core}
As stars approach the end of the TP-AGB phase, they are characterized by high luminosity ($\log L/L_\odot \gtrsim 4.2-4.4$, depending on the initial mass) and low effective temperature ($\log T_\mathrm{eff} \lesssim 3.40$). In these advanced phases, issues in finding model convergence arise and it becomes difficult to follow the evolution with \texttt{PARSEC}. 
This region of the HR diagram is notorious for numerical difficulties \citep{wood_hydrostatic_1986, wagenhuber_termination_1994, herwig_internal_1999, karakas_asymptotic_2003, miller_bertolami_full_2006, karakas_stellar_2007, weiss_new_2009, lau_end_2012}. 
These numerical difficulties appear to be independent of the specific stellar evolution code, computational grid, and time step employed (see also discussion by \citeauthor{addari_francesco_third_2020} \citeyear{addari_francesco_third_2020}). We tried to determine with our present tracks whether this challenge stems from purely numerical complications or if it originates from a more physically motivated ground; we have not found conclusive information yet, but a full analysis is out of the scope of this work.
As a preliminary note, we observed that if the star's envelope is swiftly stripped off (achieved by setting a higher mass-loss rate), the track seamlessly approaches the conclusion of the TP-AGB phase and enters the post-AGB phase.
We are aware that the mass-loss has a huge impact on TP-AGB evolution, as it sets the lifetime of this phase. Due to our assumptions on stellar winds, in particular the low mass-loss phase (region B in Figure~\ref{fig:masslossevol}), our tracks extend well toward high luminosity and low effective temperature. For this reason, our models usually do not reach the end of the TP-AGB, and the calculation stops a few pulses before the almost complete envelope ejection.

However, it is still possible to get reliable and self-consistent estimates of the final core mass. First, we employ a simple extrapolation scheme (Sect. \ref{sec:extrap}) to filter out the extreme cases that cannot reproduce the expected IFMR given by \citet{marigo_carbon_2020}. Then, we completed the remaining models with the \texttt{COLIBRI} code \citep{marigo_evolution_2013}, the characteristics of which are briefly summarized in the Sect. \ref{sec:colibri}.

\subsection{A simple estimate} \label{sec:extrap}
Before applying \texttt{COLIBRI} code to complete the tracks, we can get a general understanding of the impact of the overshooting parameters by extrapolating over the last thermal pulses. This technique exploits typical key parameters that influence the TP-AGB evolution. These quantities include: core mass $M_\mathrm{core}$, envelope mass $M_\mathrm{env}$, mass-loss rate $\dot{M}$, 3DU efficiency $\lambda$, effective temperature $T_\mathrm{eff}$ and inter-pulse period $\tau_\mathrm{int}$. The procedure is very similar to that adopted by the \texttt{COLIBRI} code but gives up much of its complexity in order to get a simple, yet effective, estimate. Indeed, the process of extrapolating or extending evolutionary models carries the inherent risk of introducing errors or inaccuracies, especially when venturing beyond simulated data points. To account for that, we introduced error bars to the results given by this extrapolation technique, by assuming parameters that intuitively give the lowest and largest core mass growth in the remaining part after \texttt{PARSEC} evolution. We extensively discuss the results of the extrapolation in Sect. \ref{sec:ifmr}.

Despite the risk of extending evolutionary models, it is recognized retrospectively that the approach undertaken serves the intended purpose effectively. Specifically, in the context of studying the IFMR, it allows us to get a first-order feeling of the impact of $(f_\mathrm{env}, f_\mathrm{pdcz})$ on the final core mass. This preliminary set of information is then used to reject some of the possibilities, reducing the set of tracks that then are properly completed with \texttt{COLIBRI}. 

\subsection{The \texttt{COLIBRI} code} \label{sec:colibri}
In comparison to fully synthetic TP-AGB codes \citep{groenewegen1993, izzard_new_2004}, the \texttt{COLIBRI} code leaves much of their analytic formalism in favor of detailed physics applied to a complete envelope model, in which the four stellar structure equations are integrated from the atmosphere to the bottom of the hydrogen-burning shell. It incorporates the \texttt{\AE SOPUS} code \citep{marigo_low-temperature_2009, marigo_low-temperature_2022} as a routine to calculate the equation of state and Rosseland mean opacities in the star's external layers. Furthermore, it accounts for hot-bottom burning nucleosynthesis and energetics using a nuclear network coupled with a diffusive treatment of convection.
The TDU is parameterized, but we recently introduced a more physically sound description in which the TDU efficiency is determined by core and envelope mass \citep{Marigo_universe_22}.

Several routines are shared by \texttt{PARSEC} and \texttt{COLIBRI}, including those related to convection, atmosphere, opacities, nuclear reactions, mass loss, and the diffusive treatment of convection in the envelope.

As a result, the \texttt{COLIBRI} code is an appropriate tool for completing the TP-AGB evolution computed by \texttt{PARSEC}.
\texttt{COLIBRI} calculations start from the last TP computed by \texttt{PARSEC}, at the quiescent stage of pre-flash luminosity maximum.
Many structural parameters are loaded into \texttt{COLIBRI} from \texttt{PARSEC} last model, including the intershell chemical composition, which will be kept constant until the end of the evolution.
\begin{figure}[!t]
    \centering
    \includegraphics[width=0.45\textwidth]{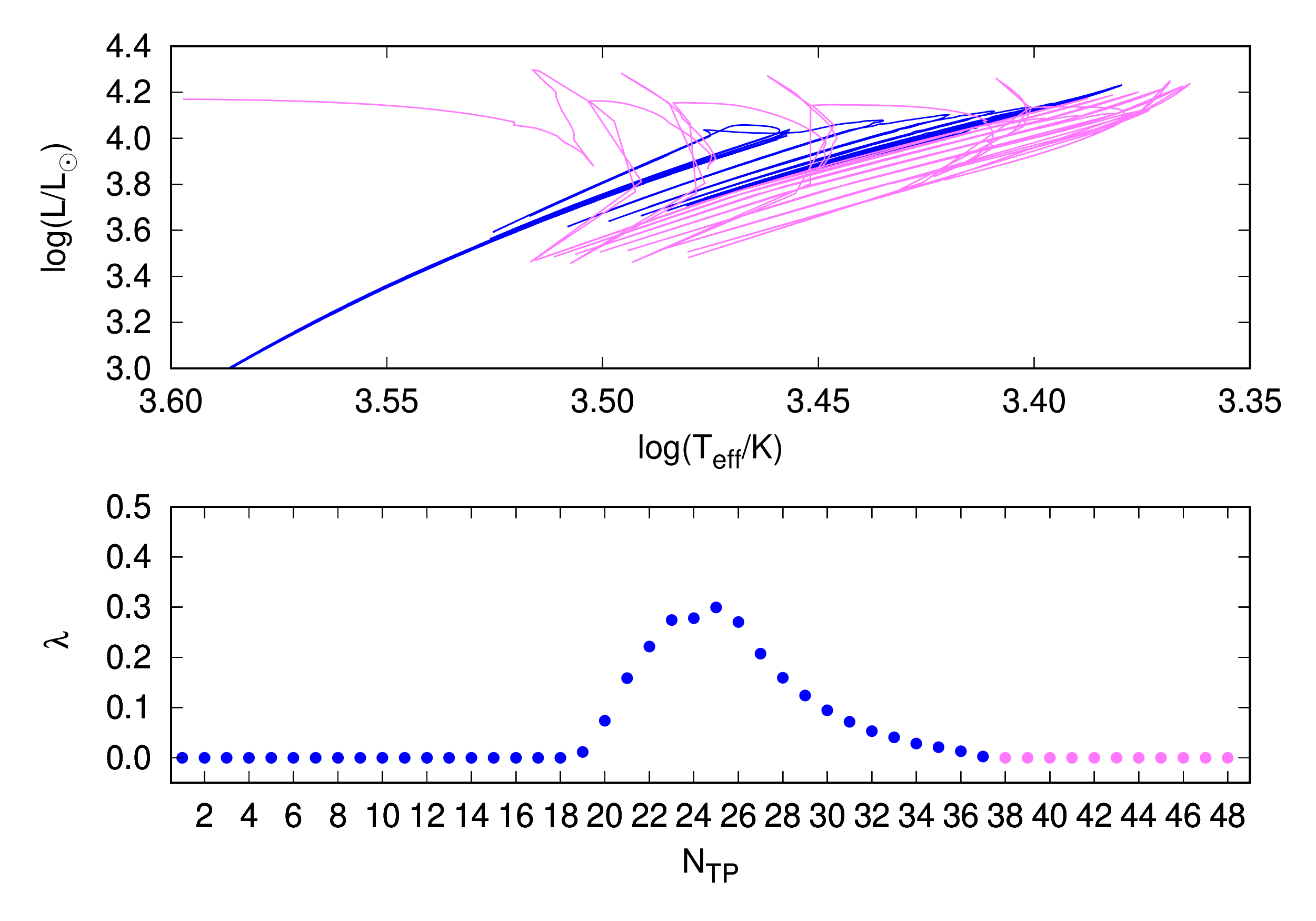}
\caption{Top panel: evolutionary track of a TP-AGB model with $\Mi=1.85\,\Msun$. \texttt{PARSEC}  and \texttt{COLIBRI} sections of the track are colored in blue and pink, respectively. Bottom panel: TDU efficiency as a function of the TP number.}
\label{fig_hr_parsec_colibri}
\end{figure}
Figure~\ref{fig_hr_parsec_colibri} shows an example of the matching between  \texttt{PARSEC} and \texttt{COLIBRI} computations. This TP-AGB model becomes a carbon star, has a final core mass of $0.726\,\Msun$, and populates the IMFR kink, in agreement with the data \citep{cummings_white_2018, marigo_carbon_2020}. We also notice that the TDU is already quenched when \texttt{COLIBRI} starts the computations. As a consequence, the number of pulses computed with \texttt{COLIBRI} ($N_\mathrm{COLIBRI}$) is modest in most tracks with  $N_{\rm COLIBRI} \simeq 1-2$, except for a few cases where $N_{\rm COLIBRI} \simeq 8-10$. We will discuss the final results with the complete tracks in the following section.

\subsection{Shape of the IFMR} \label{sec:ifmr}
\citet{marigo_carbon_2020} demonstrated that the Initial-Final Mass Relations is not
monotonic and exhibits a notable kink between initial masses around $1.65 - 2.1 M_\odot$. This kink is attributed to the chemical enrichment brought about by the third dredge-up phenomenon, coupled with the diverse regimes of mass loss as ruled by the carbon excess $\mathrm{C} - \mathrm{O}$:
\begin{equation}
    \mathrm{C} - \mathrm{O} = \log \left( \frac{n_\mathrm{C} - n_\mathrm{O}}{n_\mathrm{H}} \right) + 12 \quad \mathrm{if} \quad  n_\mathrm{C} - n_\mathrm{O} > 0
\end{equation}
where $n_\mathrm{i}$ is the number abundance of the $i$ element. Stars with carbon excess less than $8.2 \lesssim (\textrm{C} - \textrm{O})_\mathrm{min} \lesssim 9.2$ experience extended lifetimes, due to weaker winds (B-regime in Figure~\ref{fig:masslossevol}), and the core grows above the usual monotonic trend.

One advantage of employing purely synthetic or hybrid evolutionary codes is that the TDU efficiency $\lambda$ can be treated as a free parameter. The IFMR offers an excellent means to directly calibrate this efficiency without the constraints imposed by more first-principle parameters like convective overshooting efficiency. Instead, we aim to reproduce the shape of the IFMR by setting the values of ($f_\mathrm{env}$, $f_\mathrm{pdcz}$), which have a wider impact than simply changing $\lambda$.
\begin{figure*}[htbp]
    \centering
    \includegraphics[width=0.95\textwidth]{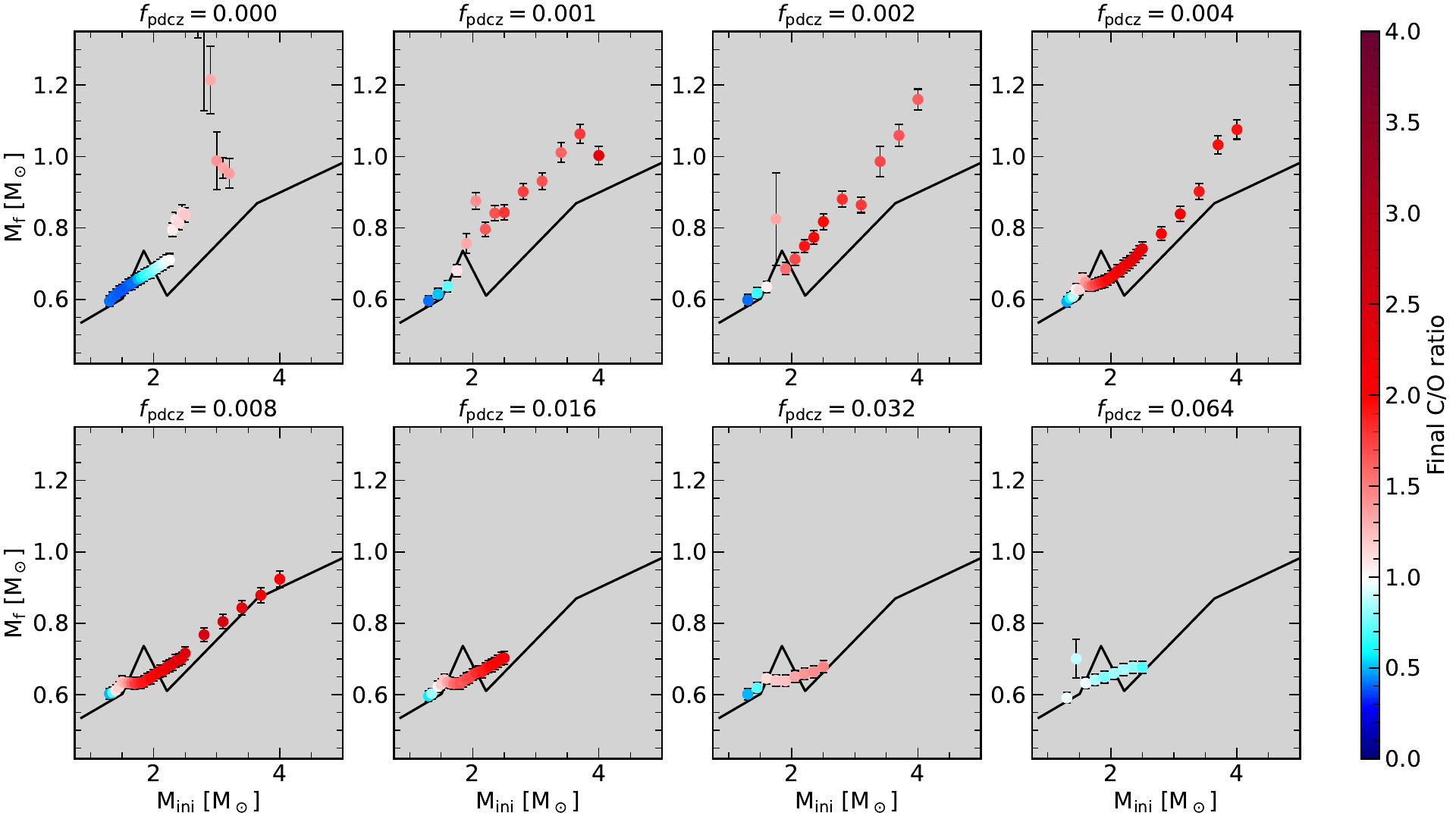}
    \caption{Final core masses for fixed envelope overshooting $f_\mathrm{env} = 0.047^\star$ and varying PDCZ overshooting. The solid line is the semi-empirical IFMR found by \citet{marigo_carbon_2020}. Sets with large overshooting efficiency are limited to $1.3 \leq M_\mathrm{ini} / M_\odot \leq 3.2$, which is sufficient to show they are not able to reproduce the IFMR's kink. The final mass is estimated with the extrapolation technique (Sect. \ref{sec:extrap}). A slight oscillation of the result is produced by the extrapolation technique (see Sect. \ref{sec:extrap} as well as the error bars.)} \label{fig:varfov_ifmr}   
\end{figure*}
\begin{figure*}[htbp]
    \centering
    \includegraphics[width=0.95\textwidth]{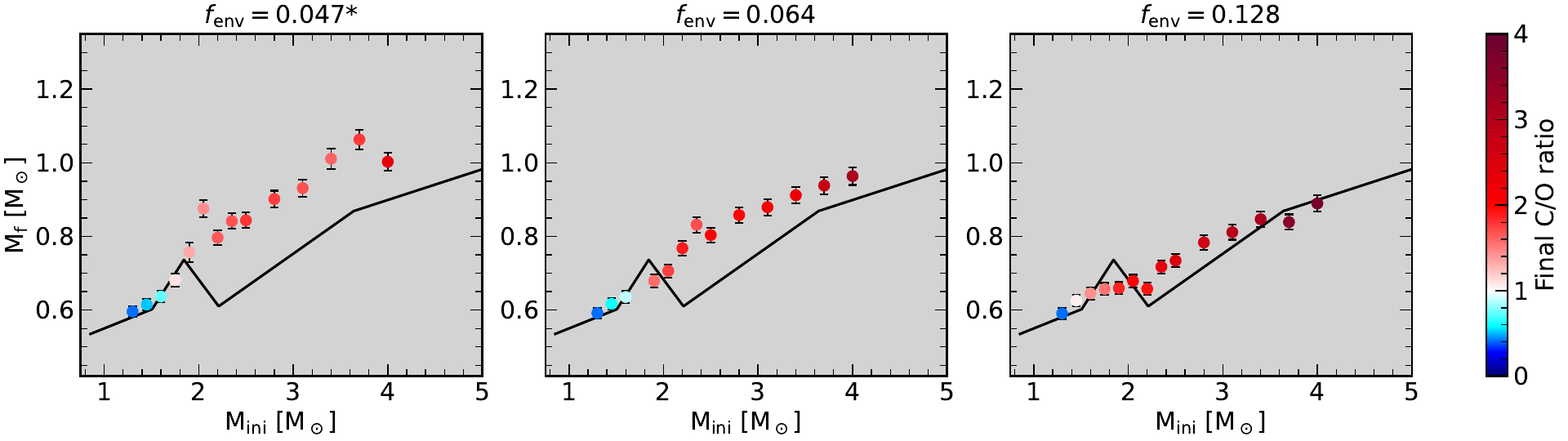}
    \caption{Final core masses for fixed PDCZ overshooting $f_\mathrm{pdcz} = 0.001$ and varying envelope overshooting. The solid line is the a fit to the semi-emipirical IFMR of \citet{marigo_carbon_2020}, \citet{Cummings_etal_18}. The final mass is estimated with the extrapolation technique (Sect. \ref{sec:extrap}).} \label{fig:varenv_ifmr}
\end{figure*}
\begin{figure*}
  \centering
  \begin{minipage}[t]{0.49\textwidth}
    \includegraphics[width=\textwidth]{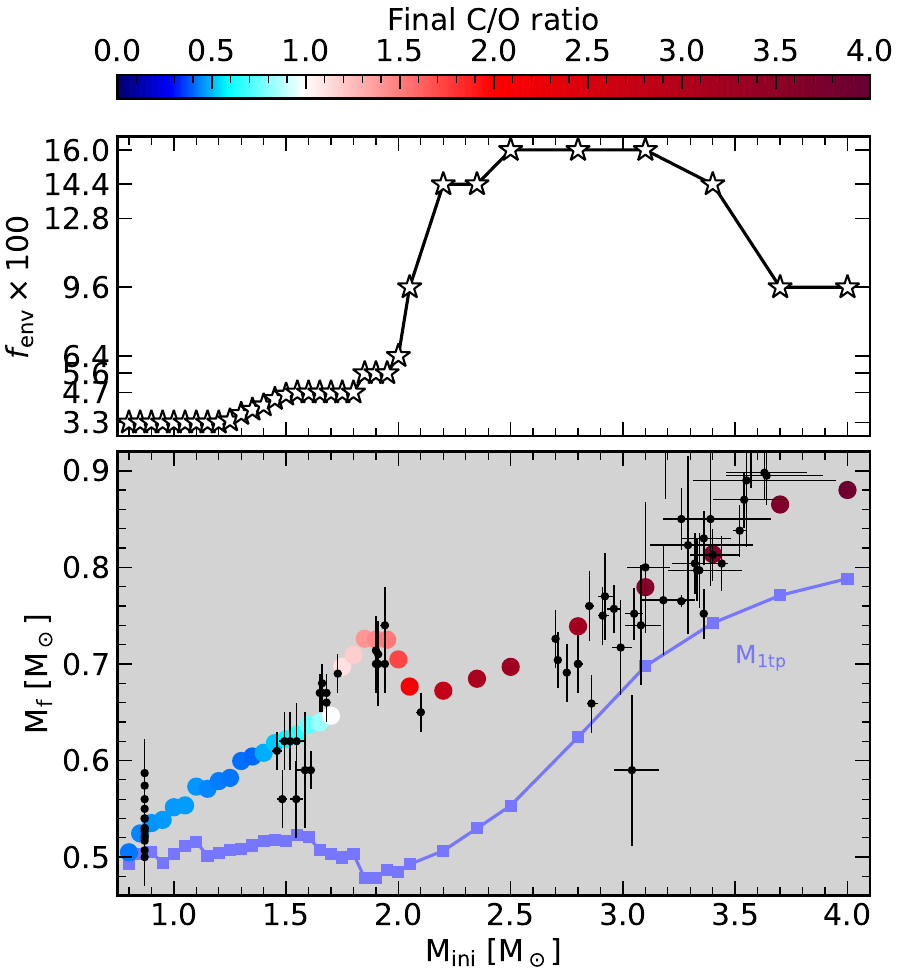} % Replace with your image file name
    \caption{Bottom panel: comparison between the semi-empirical IFMR \citep[points with error bars;][]{cummings_white_2018, marigo_carbon_2020}, and theoretical predictions of this work (filled circles) color-coded as a function of the final surface C/O. The final mass is reached with \texttt{COLIBRI} as explained in Sect. \ref{sec:colibri}. The core mass at the first thermal pulse $M_\mathrm{1tp}$ is plotted for reference. Top panel: the corresponding value of envelope overshooting, multiplied by 100 for visualization purposes. 
    }
    \label{fig:final_ifmr}
  \end{minipage}
  \hfill
  \begin{minipage}[t]{0.49\textwidth}
    \includegraphics[width=\textwidth]{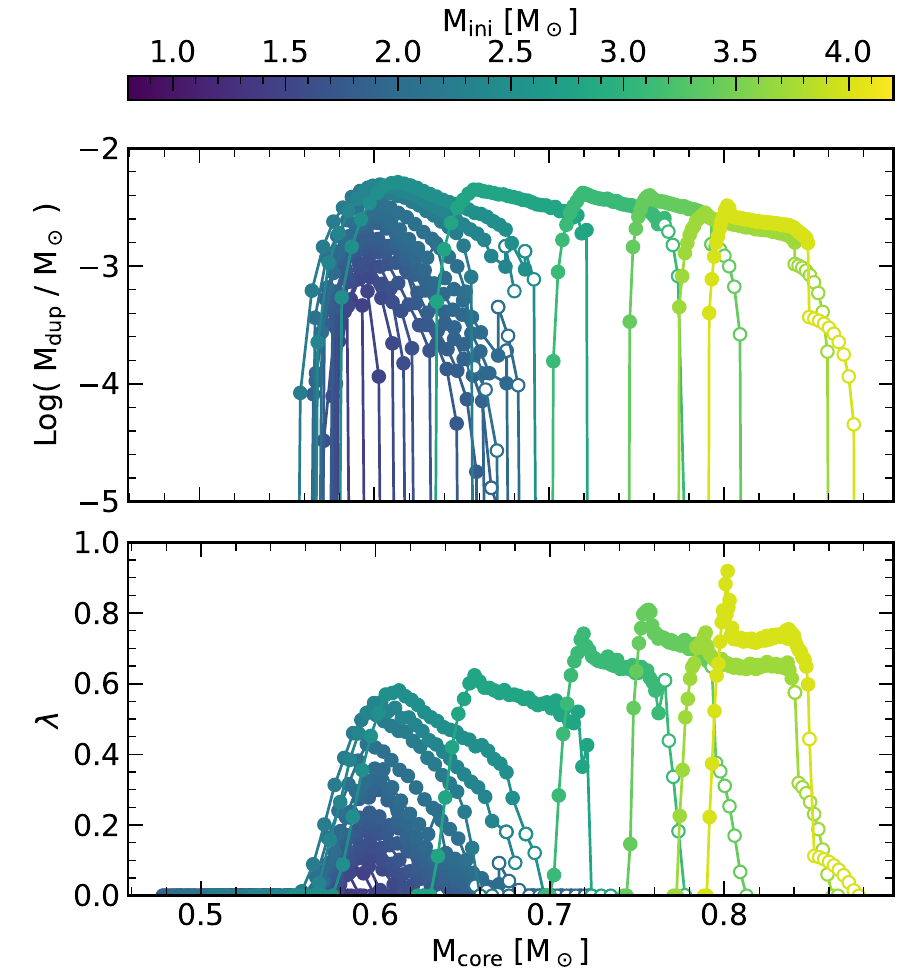} % Replace with your image file name
    \caption{Bottom panel: TDU efficiency versus current core mass for the tracks used in Figure~\ref{fig:final_ifmr}. \texttt{PARSEC} TDUs are marked by filled circles, \texttt{COLIBRI} TDU are marked with empty circles. Top panel: Log10 of the dredge-up mass versus current core mass for the same tracks and symbols.}
    \label{fig:final_ifmr_lambda}
  \end{minipage}
\end{figure*}

Figure \ref{fig:varfov_ifmr} and Figure \ref{fig:varenv_ifmr} illustrate the IFMR for varying $f_\mathrm{pdcz}$ and $f_\mathrm{env}$, respectively. In most cases, a monotonic trend is evident, without any kink. We can notice how the low mass tracks ($M_\mathrm{ini} \lesssim 1.5 M_\odot$) are close to the expected IFMR independently on the choice of the parameters. That is because these tracks experience very shallow or no dredge-up, so the core mass have similar evolution even with different ($f_\mathrm{pdcz}$, $f_\mathrm{env}$).
On the other hand, in the intermediate-mass range most of our models overestimate the final mass given by the semi-empirical IFMR. The discrepancy is progressively reduced by increasing $f_\mathrm{env}$ or $f_\mathrm{pdcz}$, meaning that a more efficient TDU is needed at these initial masses. This is consistent with the previous IFMR calibrations \citep{marigo_carbon_2020, Marigo_universe_22}, which indicate that $\lambda_\mathrm{max}$ increases with the initial mass.
Our findings show that a fixed couple of overshooting parameters for all masses cannot reproduce the kink at $M_\mathrm{ini} \simeq 1.65 - 2.10 M_\odot$.

At this point we still have two parameters that affect the shape of the IFMR. For the rest of the discussion, we fix the value of $f_\mathrm{pdcz}$ and we increase the efficiency of TDU by changing $f_\mathrm{env}$ only, thus fixing the intershell composition (see Figure~\ref{fig:intershell}).
We will discuss the robustness of this assumption later in this section. In this framework, we can immediately reject the extreme cases $f_\mathrm{pdcz} = 0.000$ and $0.064$.
The latter simply does not produce any carbon star.
The case with no overshooting in the PDCZ experiences very weak dredge-ups across the whole mass range. Then, the first carbon stars are produced at $M_\mathrm{ini} \simeq 2.30 - 2.40 M_\odot$ while we expect to find them at lower masses \citep{marigo_fresh_2021}.
Figure~\ref{fig:varfov_ifmr} shows very clearly that even small values of the PDCZ-overshooting, compared to $f_\mathrm{pdcz} = 0.000$, make the biggest impact on the IFMR. Sets with $f_\mathrm{pdcz}=0.001 - 0.002$ are already enough to accommodate all initial masses up to the kink maximum at $M_\mathrm{ini} \simeq 1.85 M_\odot$.
To select our optimal final set we select $f_\mathrm{pdcz} = 0.001$, as $f_\mathrm{pdcz} > 0.002$ produce too efficient TDU events (both in envelope penetration and carbon abundance in the intershell) already with the fiducial value ($f_\mathrm{env} = 0.047^\star$) of envelope overshooting.
Having set $f_\mathrm{pdcz}$, we can just vary $f_\mathrm{env}$ until a satisfactory fit is achieved, and we can complete these tracks with \texttt{COLIBRI} to find more accurate estimates of the final mass.
Figure~\ref{fig:final_ifmr} shows the final IFMR on top of observational data used in previous works \citep{cummings_white_2018, marigo_carbon_2020, canton_white_2021} and Figure~\ref{fig:final_ifmr_lambda} shows the impact of each dredge-up on the same set of tracks.
We collect the properties of the final set of tracks in Table~\ref{tab:env}.
\begin{table*}[htbp]
\centering
%\small
\begin{tabularx}{\textwidth}{@{\extracolsep{\fill}}ccccccccccccc}
        \toprule \toprule 
        $M_\mathrm{ini}$ & $f_\mathrm{env}$ & $N_\mathrm{TP}$ & $N_\mathrm{\texttt{COLIBRI}}$ & $M_\mathrm{f, \texttt{PARSEC}}$  & $M_\mathrm{f,\texttt{COLIBRI}}$   & (C/O)$_f$ & $\tau_\mathrm{HB}$ & $\tau_\mathrm{EAGB}$ & $\tau_\mathrm{TPAGB}$ & $\mathrm{R}_2$  & $\tau_\mathrm{C}$ & $\lambda_\mathrm{max}$ \\ 
        \midrule
    0.80 &    0.033 &        6 &         \xmark &    0.503 &      \xmark &    0.424 &  102.510 &   16.129 &    1.455 &    0.172 &    0.000 &    0.000  \\ 
    0.85 &    0.033 &        6 &         \xmark &    0.523 &      \xmark &    0.441 &  111.940 &   11.547 &    1.791 &    0.119 &    0.000 &    0.000  \\ 
    0.90 &    0.033 &        7 &         \xmark &    0.534 &      \xmark &    0.455 &  109.881 &   11.492 &    1.963 &    0.122 &    0.000 &    0.000  \\ 
    0.95 &    0.033 &       11 &         \xmark &    0.538 &      \xmark &    0.464 &  113.405 &   14.591 &    2.498 &    0.151 &    0.000 &    0.000  \\ 
    1.00 &    0.033 &       10 &         \xmark &    0.551 &      \xmark &    0.462 &  108.428 &   12.715 &    2.545 &    0.141 &    0.000 &    0.000  \\ 
    1.05 &    0.033 &        9 &         \xmark &    0.554 &      \xmark &    0.468 &  118.934 &    9.192 &    2.070 &    0.095 &    0.000 &    0.000  \\ 
    1.10 &    0.033 &       10 &         \xmark &    0.573 &      \xmark &    0.458 &  112.987 &    9.563 &    2.394 &    0.106 &    0.000 &    0.000  \\ 
    1.15 &    0.033 &       14 &         \xmark &    0.571 &      \xmark &    0.438 &  109.913 &   14.758 &    2.908 &    0.161 &    0.000 &    0.000  \\ 
    1.20 &    0.033 &       14 &         \xmark &    0.579 &      \xmark &    0.428 &  111.067 &   13.911 &    2.952 &    0.152 &    0.000 &    0.000  \\ 
    1.25 &    0.034 &       14 &         \xmark &    0.582 &      \xmark &    0.418 &  108.358 &   13.055 &    2.841 &    0.147 &    0.000 &    0.000  \\ 
    1.30 &    0.037 &       17 &        1 &    0.598 &    0.599 &    0.413 &  107.687 &   12.026 &    3.018 &    0.140 &    0.000 &    0.000  \\ 
    1.35 &    0.039 &       16 &        1 &    0.600 &    0.604 &    0.406 &  113.896 &   10.045 &    2.935 &    0.114 &    0.000 &    0.000  \\ 
    1.40 &    0.041 &       15 &        2 &    0.598 &    0.608 &    0.487 &  120.513 &    8.452 &    2.786 &    0.093 &    0.000 &    0.063  \\ 
    1.45 &    0.044 &       17 &        1 &    0.614 &    0.618 &    0.520 &  103.378 &   11.132 &    2.914 &    0.136 &    0.000 &    0.072  \\ 
    1.50 &    0.046 &       18 &        1 &    0.616 &    0.622 &    0.581 &  108.628 &   10.258 &    2.972 &    0.122 &    0.000 &    0.106  \\ 
    1.55 &    0.047 &       17 &        2 &    0.617 &    0.627 &    0.641 &  121.401 &    7.263 &    2.782 &    0.083 &    0.000 &    0.110  \\ 
    1.60 &    0.047 &       20 &        1 &    0.635 &    0.637 &    0.741 &  113.248 &    9.355 &    3.055 &    0.110 &    0.000 &    0.137  \\ 
    1.65 &    0.047 &       23 &        1 &    0.634 &    0.640 &    0.830 &  129.538 &   10.010 &    3.510 &    0.104 &    0.000 &    0.169  \\ 
    1.70 &    0.047 &       25 &        2 &    0.638 &    0.646 &    0.982 &  117.913 &   13.553 &    3.812 &    0.147 &    0.000 &    0.190  \\ 
    1.75 &    0.047 &       34 &        4 &    0.679 &    0.697 &    1.103 &  129.467 &   14.204 &    4.311 &    0.143 &    0.657 &    0.210  \\ 
    1.80 &    0.047 &       34 &        5 &    0.684 &    0.709 &    1.182 &  159.252 &    8.776 &    4.283 &    0.082 &    0.789 &    0.235  \\ 
    1.85 &    0.056 &       38 &       10 &    0.672 &    0.726 &    1.401 &  267.388 &   17.313 &    5.252 &    0.084 &    1.014 &    0.299  \\ 
    1.90 &    0.056 &       38 &       12 &    0.662 &    0.725 &    1.454 &  234.687 &   17.506 &    5.149 &    0.097 &    0.890 &    0.321  \\ 
    1.95 &    0.056 &       38 &        8 &    0.676 &    0.725 &    1.518 &  240.711 &   14.036 &    5.029 &    0.079 &    0.975 &    0.329  \\ 
    2.00 &    0.064 &       37 &        6 &    0.671 &    0.705 &    1.727 &  195.418 &   18.703 &    4.962 &    0.121 &    0.925 &    0.360  \\ 
    2.05 &    0.096 &       34 &        4 &    0.657 &    0.677 &    2.127 &  193.298 &   15.462 &    4.690 &    0.104 &    0.939 &    0.429  \\ 
    2.20 &    0.144 &       34 &        2 &    0.660 &    0.672 &    2.788 &  154.412 &   11.410 &    4.218 &    0.101 &    1.073 &    0.519  \\ 
    2.35 &    0.144 &       34 &        1 &    0.675 &    0.685 &    3.130 &  126.222 &    8.305 &    3.476 &    0.093 &    1.115 &    0.547  \\ 
    2.50 &    0.160 &       34 &        2 &    0.686 &    0.697 &    3.287 &   95.420 &    8.453 &    2.935 &    0.119 &    1.193 &    0.581  \\ 
    2.80 &    0.160 &       35 &        3 &    0.724 &    0.739 &    3.219 &   63.693 &    5.019 &    1.818 &    0.107 &    0.905 &    0.624  \\ 
    3.10 &    0.160 &       38 &        4 &    0.766 &    0.779 &    3.571 &   43.372 &    3.239 &    1.194 &    0.102 &    0.663 &    0.742  \\ 
    3.40 &    0.144 &       41 &        7 &    0.793 &    0.814 &    3.626 &   29.673 &    2.559 &    0.955 &    0.118 &    0.583 &    0.809  \\ 
    3.70 &    0.096 &       62 &       10 &    0.841 &    0.865 &    3.681 &   23.604 &    1.559 &    0.885 &    0.104 &    0.563 &    0.745  \\ 
    4.00 &    0.096 &       71 &       11 &    0.849 &    0.880 &    3.901 &   17.128 &    1.360 &    0.859 &    0.130 &    0.584 &    0.919  \\     \bottomrule \bottomrule
    \end{tabularx}
    \caption{Relevant data for the final track set. All tracks have $f_\mathrm{pdcz}=0.001$. $N_\mathrm{TP}$ is the number of pulses calculated by \texttt{PARSEC}. $N_\mathrm{\texttt{COLIBRI}}$ is the number of TPs calculated by \texttt{COLIBRI}. $M_\mathrm{f, \texttt{PARSEC}}$  is the core mass reached by \texttt{PARSEC}, $M_\mathrm{f, \texttt{COLIBRI}}$ is the final core mass calculated by continuing the evolution with \texttt{COLIBRI}. $\tau_\mathrm{HB}$ is the time spent in the core He-burning phase, $\tau_\mathrm{EAGB}$ is the time spent in Early AGB, from the central He exhaustion to the beginning of TP-AGB, set when the helium intershell thickness becomes smaller than $0.1 M_\odot$. $\tau_\mathrm{TPAGB}$ is the time spent in TP-AGB phase and $\tau_\mathrm{C}$ is the time spent as a C-type star. $\mathrm{R}_2$ ratio is estimated as $ (\tau_\mathrm{EAGB} + \tau_\mathrm{TPAGB}) / \tau_\mathrm{HB}$. Masses are in solar units and ages in Myr.}
    \label{tab:env}
\end{table*}
As expected, the negative slope region ($M_\mathrm{ini} \simeq 1.85 - 2.20 M_\odot$ ) indicates an increasing $f_\mathrm{env}$, even up to four times greater than the envelope undershooting extend in the lower mass range.
Another change of slope is evident at approximately $M_\mathrm{ini} \sim 3.50 M_\odot$, where a reduction in $f_\mathrm{env}$ is necessary to accommodate larger core growth. 
In this high-mass range, core growth is severely hampered by the combination of high-efficiency TDU and hot bottom burning (HBB), significantly limiting core growth during the interpulse phase.
However, the white dwarf data spread for $M_\mathrm{ini} \simeq 3.00 M_\odot$ give looser constraints on the $f_\mathrm{env}$ and we selected an average value of $f_\mathrm{env}$.
Figure~\ref{fig:final_ifmr_lambda} shows again that \texttt{PARSEC} calculation stops close to the end of TP-AGB, where TDU is quenched. Then, it is clear that \texttt{COLIBRI} has a limited impact on the results. 
Nevertheless, it stands out as a very efficient and powerful tool to complete the TP-AGB evolution, where full stellar evolutionary codes usually give up.

Finally, we want to briefly discuss the robustness of the results giving up the assumption of fixed $f_\mathrm{pdcz}$.
If we let both parameters free, a simple $\mathrm{\chi}^2$-fit gives a random distribution of ($f_\mathrm{env}, f_\mathrm{pdcz}$), with no clear trend with the initial mass.
Still leaving it as an open possibility, it would give little or no information on the internal mixing processes and on the efficiency of the third dredge-up.
That is because the parameters not only change the envelope penetration but also the intershell composition, thus leading to a mild degeneracy of the parameters.
The works by \citet{wagstaff_impact_2020} and \citet{addari_francesco_third_2020} give useful hints on how to overcome this apparent degeneracy. The [WC]-type of central stars of planetary nebula abundances reflects the intershell composition, giving tighter constraints on the $f_\mathrm{pdcz}$ only.
These findings suggested us to select a value for $f_\mathrm{pdcz}$, and let only $f_\mathrm{env}$ vary to adjust the efficiency of TDU. \citet{Herwig_2005} and \citet{wagstaff_impact_2020} suggest $0.004 < f_\mathrm{pdcz} < 0.016$, which is much higher than $f_\mathrm{pdcz} = 0.001$ found in this work based on IFMR. However, the typical errors for PG 1159 stars abundances (about 0.3-0.5 dex, \citeauthor{werner_elemental_2006} \citeyear{werner_elemental_2006}; \citeauthor{werner_weak_2014} \citeyear{werner_weak_2014}; \citeauthor{werner_far-ultraviolet_2016} \citeyear{werner_far-ultraviolet_2016}) and the data spread make it difficult to differentiate between scenarios with low or zero $f_\mathrm{pdcz}$, by relying solely on intershell abundances.
Given the intricacies in interpreting observations, primarily due to the potential influence of late thermal pulses, the study of [WC]-type of central stars of planetary nebula is beyond the scope of this work, but we acknowledge that coupling this data with the IFMR can potentially lift the degeneracy of $(f_\mathrm{env}, f_\mathrm{pdcz})$.

\section{Concluding remarks} \label{sec:conclusion}

This study undertook a comprehensive exploration of a broad spectrum of values for both envelope and pulse-driven convective zone overshooting, aiming to calibrate them in accordance with the semi-empirical Initial-Final Mass Relation presented by \citet{marigo_carbon_2020}. Notably, this work stands as the first successful endeavor in reproducing the IFMR kink employing a full stellar evolution code, \texttt{PARSEC}, supported by the \texttt{COLIBRI} code. Given that \texttt{PARSEC} stops in close proximity to the end of their evolution, where the third dredge-up is no longer relevant in shaping the core mass, and considering the inherent scatter in observational data, the validity of our scheme is evident.

It is worth noting that the previous TDU calibration based on the semi-empirical IFMR performed by \citet{marigo_carbon_2020} and \citet{Marigo_universe_22} is supported by this work. White dwarfs populating the kink had progenitor carbon stars that experienced shallow mixing events ($\lambda_{\rm max} \lesssim 0.2$) with mild carbon enrichment and modest mass-loss rates. Then, the efficiency of the TDU increases with increasing initial mass in both \texttt{PARSEC} and \texttt{COLIBRI} codes.
This suggests that the efficient computational capabilities of the \texttt{COLIBRI} code make it a valuable tool for both exploring the parameter space and potentially indicating a calibration pathway.

In this study, we show that the fiducial value of envelope overshooting ($f_\mathrm{env} = 0.047^\star$) alone falls short of explaining the observed outcomes, as such tracks dredge up minimal amounts of mass.
Our investigation establishes that a constant envelope penetration is not a viable solution either.
On the other hand, a small value for the PDCZ overshooting ($f_\mathrm{pdcz}=0.001$) with a varying $f_\mathrm{env}$ (Figure~\ref{fig:final_ifmr}, top panel) proves sufficient to approach a feasible solution in reproducing the IFMR. This work may serve as a practical guideline, focusing on the extent of the envelope penetration (Figure~\ref{fig:final_ifmr_lambda}) and the composition of the intershell (Fig.~\ref{fig:intershell}, panel $f_\mathrm{pdcz}=0.001$). The cumulative effects of these two factors directly influence mass loss, ultimately shaping the star's evolution to align with the observed IFMR. However, we noted a mild degree of degeneracy between the two overshooting parameters. Our future exploration will go deeper into this aspect, by combining the present results with information from observed intershell composition from [WC]-type central stars of planetary nebulae \citep{wagstaff_impact_2020}. This extension holds the potential to constrain the value of $f_\mathrm{pdcz}$ independently, thus leaving only the envelope overshooting efficiency as a free parameter.

%% Also note that the akcnowlodgment environment does not support long amounts of text. If you have a lot of people and institutions to acknowledge, do not use this command. Instead, create a new \section{Acknowledgments}.
\begin{acknowledgments}
Acknowledgments: PM, FA, AB, GV acknowledge funding support by the Italian Ministerial Grant PRIN 2022, “Radiative opacities for astrophysical applications”, no. 2022NEXMP8, CUP C53D23001220006. GC acknowledges support from the Agence Nationale de la Recherche grant POPSYCLE number ANR-19-CE31-0022.
We thank the anonymous referee for providing valuable comments that helped to improve this manuscript.
\end{acknowledgments}





\appendix

\section{Shell shifting} \label{sec:shellshift}
The concept of shell shifting was initially introduced by \citet{weigert_sternentwicklung_1966}.
This idea stems directly from the principle of energy conservation.
The variation in the hydrogen profile during an interpulse phase can be expressed as follows:
\begin{equation} \label{eq:shellshift}
    \frac{\partial X}{ \partial M} \left( \frac{L_\mathrm{shell}}{X_\mathrm{env} Q} \right) = \frac{\varepsilon}{Q}
\end{equation}
where $X$ is the hydrogen abundance (being $X_\mathrm{env}$ the one in the convective envelope), $L_\mathrm{shell}$ is the hydrogen burning luminosity generated in the H-shell, $Q$ is the effective q-value of the nuclear network, $\epsilon$ is the nuclear energy generation rate and finally $M$ is the mass coordinate.
Equation \ref{eq:shellshift} was used to determine the hydrogen profile following the expansion of the core by a certain amount.
This approach demonstrated good agreement with calculations using small time steps, albeit with the caveat that computational power in the 1960s was significantly more limited than today's standards.
Nevertheless, contemporary stellar evolution codes have evolved to become more comprehensive, and TP-AGB models, in particular, remain computationally intensive.

\citet{stancliffe_third_2005} studied the difference in using simultaneous and non-simultaneous solvers in stellar evolution codes.
In conclusion, the only difference is that the second type of codes only need shorter timesteps to produce the same results, and there is no inherent problem with using a non-simultaneous method of solution.
However, shorter timesteps increase considerably computational time.
To address this issue, we have introduced a modified shell-shifting procedure to use during interpulse periods.
Diverging from the original method employed by \citet{weigert_sternentwicklung_1966}, our approach involves applying the same principle to refine the solution provided by the chemical module solver.
Instead of keeping the structure fixed while solving the nuclear network, we adjust the temperature and density profiles based on the observed shifts in the hydrogen profile.
This innovation permits the use of considerably longer time steps, extending from the range of 1-10 years to 100-1000 years.
Additionally, we introduced an energy conservation validation mechanism, which is based on an enhanced version of Eq. \ref{eq:shellshift}.
This validation procedure considers both the hydrogen abundance at the shell's upper boundary and the collective effective q-value of all hydrogen-burning reactions.
This validation step refines the final time step allocated for the chemical solution by introducing a corrective factor that ensures Eq. \ref{eq:shellshift} is respected.

This refined approach substantially reduces computational time, approximately by a third compared to the case of utilizing small time steps, while incurring minimal to no loss in accuracy concerning core growth and energy conservation.

\begin{figure}[!t]
    \centering
    \includegraphics[width=0.99\textwidth]{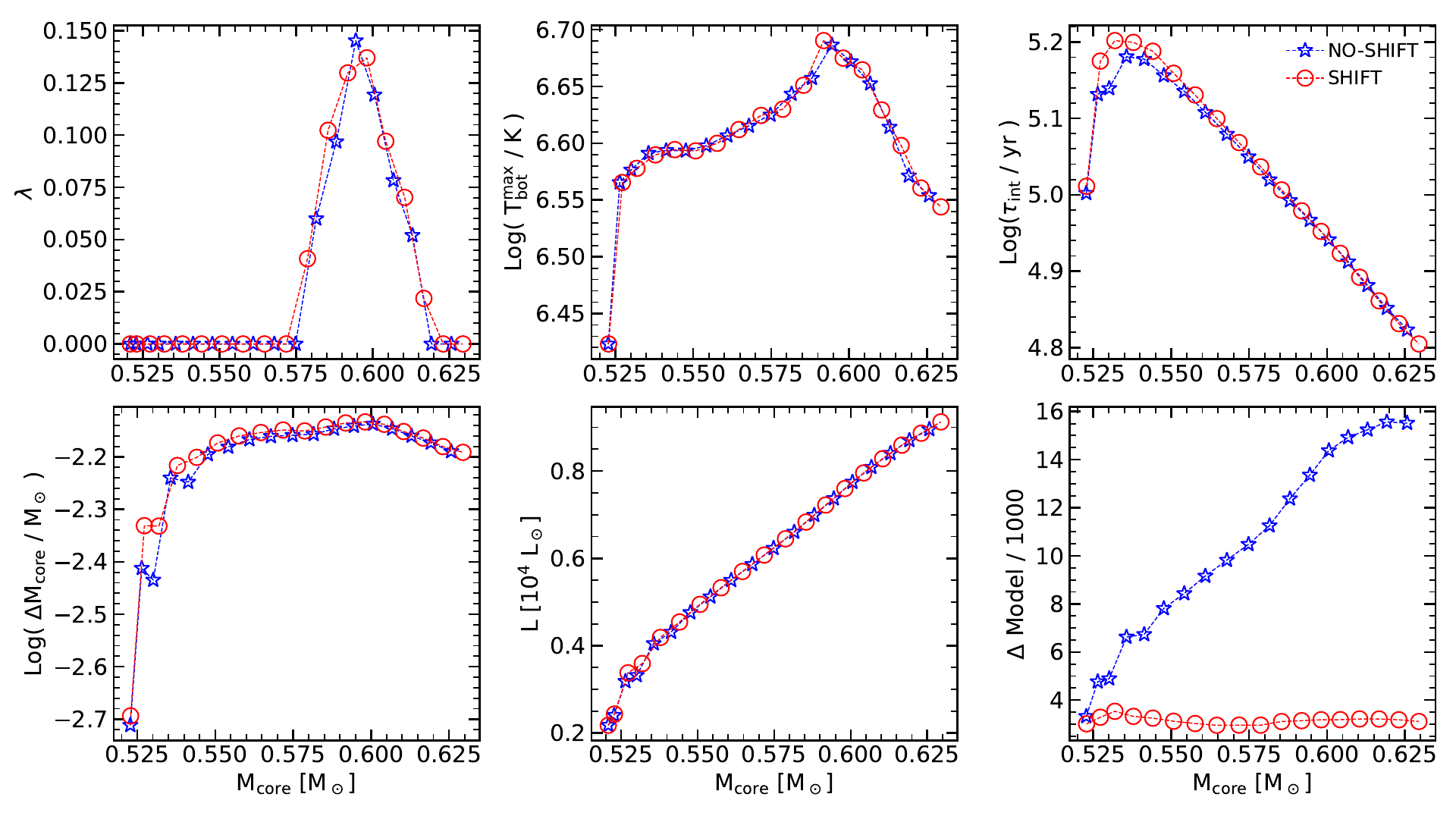}
    \caption{Comparison of the same track $M_\mathrm{ini} = 1.60 M_\odot$ from the $(0.047,0.001)$ set computed with shell-shifting (\textit{SHIFT}) and no shell-shifting (\textit{NO-SHIFT}). Top row, from left to right: TDU efficiency, temperature maximum at the bottom of the convective envelope, and interpulse time against core mass. Bottom row, from left to right: core growth per pulse cycle, maximum quiescent luminosity, and number of models per pulse cycle against core mass.}
    \label{fig:shell-shift}
\end{figure}

We computed a test track with and without the shell-shifting approach, not changing any other physical or numerical input, and we summarised the result in Figure~\ref{fig:shell-shift}. All physical quantities are in good agreement between the two cases, with only minimal differences in the first few pulses. With a minimal accuracy price, the method can speed up computing time and save storage. In the last panel of Figure~\ref{fig:shell-shift}, the number of models per pulse cycle is shown against the core mass. The shell-shifting track keeps a constant number of models (about 4000 timestamps) against a steeply rising trend topping at about 16000 models, gaining a factor of $\sim$3-4 in CPU time.

\bibliography{references.bib}{}
\bibliographystyle{aasjournal}



\end{document}